\documentclass[aip,pop,reprint]{revtex4-1}
\usepackage{graphicx}
\usepackage{hyperref}
\usepackage{amsmath}
\usepackage{stackengine}
\usepackage{ulem}
\newcommand{\tderiv}[1]{\frac{\partial #1}{\partial t}}
\newcommand{\ttderiv}[1]{\frac{\partial^2 #1}{\partial t^2}}
\newcommand{\dotcross}{{\tiny\Shortstack[c]{$\bullet$ $\times$}}}
\newcommand{\pgrad}{\nabla^\perp}
\newcommand{\llgrad}{\nabla^\parallel}
\newcommand{\pLap}{\Delta^\perp}
\newcommand{\llLap}{\Delta^\parallel}
\newcommand{\psderiv}{\partial_{\psi_v}}
\newcommand{\llderiv}{\partial^\parallel}
\newcommand{\const}{\mathrm{const}}
\newcommand{\st}[1]{\hbox{\sout{$#1$}}}

\draft 

\begin{document}


\title{A three-dimensional reduced MHD model consistent with full MHD} 



\author{N. Nikulsin}
\author{M. Hoelzl}
\affiliation{Max Planck Institute for Plasma Physics, Boltzmannstr. 2, 85748 Garching, Germany}
\author{A. Zocco}
\affiliation{Max Planck Institute for Plasma Physics, Wendelsteinstr. 1, 17491 Greifswald, Germany}
\author{K. Lackner}
\author{S. G\"{u}nter}
\affiliation{Max Planck Institute for Plasma Physics, Boltzmannstr. 2, 85748 Garching, Germany}


\date{\today}

\begin{abstract}
Within the context of a viscoresistive magnetohydrodynamic (MHD) model with anisotropic heat transport and cross-field mass diffusion, we introduce novel three-term representations for the magnetic field (background vacuum field, field line bending and field compression) and velocity (\(\vec E\times\vec B\) flow, field-aligned flow and fluid compression), which are amenable to three-dimensional treatment. Once the representations are inserted into the MHD equations, appropriate projection operators are applied to Faraday’s law and the Navier-Stokes equation to obtain a system of scalar equations that is closed by the continuity and energy equations. If the background vacuum field is sufficiently strong and the \(\beta\) is low, MHD waves are approximately separated by the three terms in the velocity representation, with each term containing a specific wave. Thus, by setting the appropriate term to zero, we eliminate fast magnetosonic waves, obtaining a reduced MHD model. We also show that the other two velocity terms do not compress the magnetic field, which allows us to set the field compression term to zero within the same reduced model. Dropping also the field-aligned flow, a further simplified model is obtained, leading to a fully consistent hierarchy of reduced and full MHD models for 3D plasma configurations. Finally, we discuss the conservation properties and derive the conditions under which the reduction approximation is valid. We also show that by using an ordering approach, reduced MHD equations similar to what we got from the ansatz approach can be obtained by means of a physics-based asymptotic expansion.
\end{abstract}


\maketitle 

\section{Introduction}

With the development of more powerful computers in the 1990s and 2000s, nonlinear numerical simulations began to play an increasingly important role in the interpretation of experimental results, planning of new experiments and the design of new machines. Nonlinear simulations allow one to simulate the operation of an entire machine on short time scales, typically thousands to hundreds of thousands of Alfv\'{e}n times. In order to simulate such time scales with a reasonable spatial resolution and using a reasonable amount of computer time, one most often has to employ reduced MHD models, which eliminate fast magnetosonic waves while retaining the relevant physics \cite{strauss1997reduced,jardin2012multiple}. The removal of fast magnetosonic waves, the fastest waves in the system, allows one to use larger time steps due to the Courant condition. Even when implicit time integration methods are used, and the Courant condition is no longer a hard limit, using time steps that are large compared to the shortest time scale can lead to poor accuracy \cite{jardin2012multiple,kruger1998generalized}. In addition, reduced MHD has less unknowns compared to full MHD, which decreases the computational costs and memory requirements for simulations.

Reduced MHD, as first introduced by Greene and Johnson \cite{greene1961determination}, and later developed by Kadomtsev, Pogutse and Strauss \cite{kadomtsev1973nonlinear,strauss1976nonlinear}, relies on ordering in a small parameter, often taken to be the inverse aspect ratio. The ordering itself is a system of several approximations and assumptions involving the ordering parameter that allows one to determine the relative order (in terms of the ordering parameter) of any quantity with respect to any other quantity of the same dimension. In this context, terms corresponding to fast magnetosonic waves have a higher order than the terms that one wants to keep, allowing the fast wave terms to be dropped. Naturally, there are many choices one can make in the ordering assumptions, depending on which physical effects one wants to keep, all of which result in different reduced equations \cite{strauss1997reduced,strauss1976nonlinear,strauss1977dynamics,strauss1980stellarator,kruger1998generalized}. The ideas of reduced MHD have also found use in astrophysics, where toroidal geometry cannot be assumed, and thus the inverse aspect ratio cannot be used as an ordering parameter \cite{oughton2017reduced}.

Starting in the 1980s, a new ansatz-based approach was introduced by Park \emph{et al} \cite{park1980non}, where an ansatz form that eliminates fast magnetosonic waves is used for the velocity and terms of all orders are kept. Izzo \emph{et al} used a similar ansatz in their study \cite{izzo1985reduced}. Later papers also adopt an ansatz for the magnetic field that eliminates field compression \cite{breslau2009some,franck2015energy}. The ansatz approach allows one to make less assumptions (although, as we will show, an assumption on the relative magnitude of the induced magnetic field still needs to be made to eliminate fast magnetosonic waves) and keep more physical effects, while generally resulting in more complicated equations than the ordering approach. For example, internal kink modes cannot be modelled at the lowest order in inverse aspect ratio, but ansatz-based models overcome that by not neglecting higher order terms \cite{park1980non}. An ansatz-based approach can also guarantee exact conservation of energy, as in the models that we derive in this paper. Thus, while keeping more physics, the various terms in the equations of ansatz-based reduced MHD are harder to interpret due to their complexity. In addition, without an ordering parameter, error estimation becomes much more difficult.

Most of the previously published versions of reduced MHD focused on tokamaks, expanding the magnetic field around a tokamak-like vacuum field, \(\vec B_v = F_0\nabla\phi\). In addition, most of the reduced models for stellarators only approximately satisfy the condition \(\nabla\cdot\vec B = 0\). A notable exception is the model by Strauss \cite{strauss1997reduced}, which expands the magnetic field around an arbitrary vacuum field, thus making no assumptions about the underlying geometry, while exactly satisfying the divergence-free condition. In this paper, we systematically follow the ansatz approach, proving our claims more rigorously than previous publicatios have. We arrive at a hierarchy of models that is mostly a generalization (with some modifications) of the results of Breslau \emph{et al} \cite{breslau2009some}, while being compatible with an arbitrary vacuum field, just as Strauss's model \cite{strauss1997reduced}. The consistency with an ordering approach is also discussed.

The remainder of the paper is structured as follows. Section \ref{sec:background} introduces the specific viscoresistive MHD equations that serve as a starting point for the derivation. The same section also introduces Clebsch-type coordinates and our representations for magnetic field and velocity. In section \ref{sec:waves}, we show that if the background vacuum field is sufficiently strong and the \(\beta\) is low, MHD waves are approximately separated by the three terms in the velocity representation, with each term containing a specific wave. In section \ref{sec:derivation}, we derive a set of scalar full MHD equations by inserting the representations into the viscoresistive MHD equations and projecting the vector equations. In section \ref{sec:rmhd}, we reduce our equations by dropping the term corresponding to fast magnetosonic waves from the velocity representation. Section \ref{sec:conservation} considers the local conservation properties of the reduced MHD equations and derives validity conditions for the reduction. Finally, in appendix \ref{sec:ordering} we show how a similar, though not identical, system of equations can be derived using an ordering approach.

\section{Viscoresistive MHD and mathematical background}
\label{sec:background}

Ideal MHD, which assumes that the plasma is a perfectly conducting inviscid fluid and that there are no sources or sinks in any of the equations, remains the most well studied plasma fluid model. Due to its simplicity, it is often used in analytical calculations, and is commonly presented in introductory texts. Despite that, non-ideal effects, such as tearing modes and other resistive instabilities, become important on longer time scales, and an equilibrium that is ideally stable may not actually be stable. Thus, most modern fluid codes employ viscoresistive (and often extended) MHD, which includes non-ideal terms, making the model more realistic at the expense of increasing equation complexity.

\subsection{Viscoresistive MHD with heat conduction}
\label{sec:vrmhd}

In this subsection, the full MHD model is introduced in its usual formulation, and will be recast using potentials and stream functions in section \ref{sec:derivation}. The usual MHD notation is followed with \(\rho\), \(p\), \(\vec v\) and \(\vec B\) being density, pressure, velocity and magnetic field, respectively. In addition to that, \(\eta\) is the resistivity, \(\nu\) is the kinematic viscosity, \(D_\perp\) is the mass diffusion coefficient perpendicular to field lines, \(\kappa_\perp\) and \(\kappa_\parallel\) are the thermal conductivity across and along field lines, and \(S_\rho\) and \(S_e\) are source terms in the continuity and energy equations, respectively. The ideal gas law \(p = \rho RT\) is assumed to hold.
\begin{widetext}
\begin{equation}
\label{eq:vrmhd}
\begin{gathered}
\tderiv{\rho} + \nabla\cdot (\rho\vec v) = P, \\
\tderiv{} (\rho\vec v) + \nabla\cdot (\rho\vec v\vec v) = \vec j\times\vec B - \nabla p + \rho\nu\Delta\vec v, \\
\tderiv{} \left(\frac{\rho v^2}{2} + \frac{p}{\gamma-1} + \frac{B^2}{2\mu_0}\right) + \nabla\cdot\left[\left(\frac{\rho v^2}{2} + \frac{\gamma p}{\gamma-1}\right)\vec v + \frac{p}{\gamma-1}\frac{D_\perp}{\rho}\nabla_\perp\rho + \frac{\vec E\times\vec B}{\mu_0} - \kappa_\perp \nabla_\perp T - \kappa_\parallel \nabla_\parallel T\right] = S_e - \frac{v^2}{2}P, \\
\tderiv{\vec B} = -\nabla\times\vec E, \\
\begin{aligned}
\nabla\times\vec B &= \mu_0 \vec j, & \nabla\cdot\vec B &= 0, & \vec E &= -\vec v\times\vec B + \eta\vec j, & P = \nabla\cdot (D_\perp \nabla_\perp \rho) + S_\rho.
\end{aligned}
\end{gathered}
\end{equation}
\end{widetext}
The gradient operators parallel and perpendicular to the total magnetic field \(\vec B\) are defined as \(\nabla_\parallel = \frac{\vec B}{B^2}\vec B\cdot\nabla\) and \(\nabla_\perp = \nabla - \nabla_\parallel\). The electric field can be eliminated from Faraday's law by inserting Ohm's law, the resulting equation will be referred to as the induction equation throughout the rest of this paper.

Note the form of the viscosity term in the Navier-Stokes equation. This is a rather simple approximation for the divergence of the viscous stress tensor in a plasma, made under the assumption that the viscosity is isotropic. While this may not be accurate for a magnetized plasma, it can nevertheless be sufficient to satisfactorily model plasma behavior \cite{huysmans2007mhd,huysmans2009non,pamela2017recent}, and including a viscosity term can sometimes help against numerical instabilities \cite{franck2015energy}. Due to the generic form of the viscosity term, it will not be treated in the derivations that follow, and instead a generic viscosity term will be added to the final equations.

\subsection{Clebsch-type coordinates}

As shown in \cite{d2012flux,stern1970euler}, any magnetic field can be locally represented in the Clebsch form:
\begin{equation}
\label{eq:clebsch}
\vec B = \nabla\alpha\times\nabla\beta.
\end{equation}
This means that the variables \(\alpha\) and \(\beta\) can be locally used to label field lines, and can parameterize any surface that the field lines intersect. A third coordinate that measures in the direction of the field is needed for a three-dimensional coordinate system. Let \(\chi\) be the magnetic scalar potential, so that \(\vec B_v = \nabla\chi\) is the vacuum component of the magnetic field, and \(\vec B - \nabla\chi\) is the induced magnetic field generated by currents flowing in the plasma. Assuming that such a background vacuum field exists, the magnetic scalar potential can be used as a third coordinate, forming the curvilinear coordinate system \((\alpha,\beta,\chi)\). Coordinate systems that rely on \(\alpha\) and \(\beta\) as the first two coordinates are called Clebsch-type coordintes, and are in general non-orthogonal \cite{d2012flux}, with orthogonal coordinate systems being possible only in the special case of a shear-free magnetic field \cite{salat2000conditions}.

Using the formalism of general curvilinear coordinates, one can define the contravariant and covariant basis vectors as:
\begin{equation}
\label{eq:basisvecs}
\begin{aligned}
&\vec e~^\alpha \equiv \nabla\alpha, & &\vec e~^\beta \equiv \nabla\beta, & &\vec e~^\chi \equiv \nabla\chi, \\
&\vec e_\alpha \equiv J\nabla\beta\times\nabla\chi, & &\vec e_\beta \equiv J\nabla\chi\times\nabla\alpha, & &\vec e_\chi \equiv J\nabla\alpha\times\nabla\beta.
\end{aligned}
\end{equation}
Covariant and contravariant components of vectors will be represented by subscripts and superscripts, respectively. Note that the third covariant basis vector points in the direction of the magnetic field. The Jacobian for this coordinate system is \(J = [\nabla\chi\cdot(\nabla\alpha\times\nabla\beta)]^{-1} = (\nabla\chi\cdot\vec B)^{-1} = 1/B^\chi\). Also note that the vacuum field will be assumed to be static in time, whereas the total magnetic field can vary. Thus, while in many similarly derived coordinate systems (see for example \cite{d2012flux}) it is assumed that \(B_\chi \equiv 1\), this cannot hold in the most general case which we will consider.

We will make use of two different Clebsch-type coordinate systems in this paper. The first system is aligned to the total magnetic field and is defined as above with \(\vec B\) being the total magnetic field, while for the second system we use the vacuum field \(\vec B_v = \nabla\chi = \nabla\psi_v\times\nabla\beta_v\) in place of \(\vec B\), aligning it to the vacuum field. To distinguish between the two systems, we label, in the vacuum field-aligned system, the first two coordintes as \(\psi_v\) and \(\beta_v\) and the basis vectors with the letter \(b\) instead of \(e\). The third coordinate \(\chi\) is shared by both systems. When flux surfaces exist, it is convenient to choose \(\psi_v\) to be the flux surface label, however none of the results of this paper depend on the choice of \(\psi_v\). For the vacuum field-aligned system, we have \(J = 1/B_v^2\) and the \(\chi\) basis vectors are related by \(\vec b_\chi = J\vec b^\chi\). Thus, \(\nabla\chi\) is perpendicular to both \(\nabla\psi_v\) and \(\nabla\beta_v\). This simplifies the coordinate system somewhat, as we have for some of the off-diagonal components of the metric tensor: \(g^{\psi_v\chi} = g^{\beta_v\chi} = g_{\psi_v\chi} = g_{\beta_v\chi} = 0\), however the off-diagonal component \(g^{\psi_v\beta_v}\) is nonzero, and the coordinate system is still non-orthogonal. The off-diagonal components are related as \(g_{\psi_v\beta_v} = -Jg^{\psi_v\beta_v}\). This becomes obvious when we recognize \(g^{ik}\) as the inverse matrix of \(g_{ik}\) and \(J\) as the determinant of \(g_{ik}\). Given that the metric tensor is symmetric and \(g_{\psi_v\beta_v}\) is the only nonzero off-diagonal component, we can immediately see that \(g_{\psi_v\beta_v} = -Jg^{\psi_v\beta_v}\).

We point out that satisfying the Laplace equation is the only requirement for \(\chi\). While in this paper we refer to \(\nabla\chi\) as the vacuum field, nothing is stopping us from setting \(\chi = F_0\phi\), where \(\phi\) is the toroidal angle, as is often done in reduced MHD for tokamaks. In that case, \(\nabla\chi\) would correspond to the dominant toroidal component of the vacuum field, while the field generated by poloidal coils will be grouped with the induced field. The magnetic scalar potential can in general be a multi-valued function. However, in the case of toroidal systems, such as tokamaks or stellarators, \(\chi\) is known analytically as a series of harmonics for arbitrary configurations, including stochastic field regions \cite{dommaschk1986representations,thomas1997new}, and the multi-valuedness is entirely contained in a tokamak-like \(F_0\phi\) term. To determine the coefficients of the toroidal and poloidal harmonics in the series, one needs to solve a system of linear algebraic equations, which fits the field \(\nabla\chi\) to a known curl-free field, usually obtained from the Biot-Savart law for the coils \cite{thomas1997new}.

It is important to note that in general the expression \eqref{eq:clebsch} is only valid locally within the vicinity of a given point. If we try to apply it globally, the functions \(\alpha\) and \(\beta\) can become multi-valued \cite{stern1970euler}. Consider, for example, the case of a tokamak with a stochastic magnetic field in a certain volume \cite{wolf2005effect}, where the same field line circles the torus infinitely many times and covers all points within the volume. Since \(\alpha\) and \(\beta\) must be constant along a field line, if we require them to be single-valued, they will be constant within the volume, leading to zero gradients and forcing the field, as given by \eqref{eq:clebsch}, to be zero, a contradiction. However, \(\alpha\) and \(\beta\) can be defined locally, starting in a given poloidal plane, parameterizing all points in the plane with coordinates \(\alpha\) and \(\beta\) (Note that once \(\alpha\) is selected, \(\beta\) has to be chosen appropriately to avoid having a scalar factor in front of the cross product in \eqref{eq:clebsch}, see \cite{d2012flux,stern1970euler}.), labeling the field lines that pass through point \((\alpha,\beta)\) with the same \(\alpha\) and \(\beta\), and then following the field lines. As long as only a sector of the torus is considered, all coordinates can be single-valued, but once one transit around the torus is made, the field lines will encounter previously labeled points. This situation is similar to what happens in coordinate systems with angular coordinates, and the solution is the same: we introduce a cut on which the \(\alpha\) and \(\beta\) undergo a discontinuity to avoid multi-valuedness, similarly to how most angular coordinates undergo a discontinuity and return to 0 after reaching \(2\pi\).

As will be shown, only \(\chi\) and \(\psi_v\) appear in the equations that we derive. Thus, in practice, it is only necessary to implement \(\chi\) and \(\psi_v\) in a code. While \(\chi\) is known analytically as a series of harmonics and only the coefficients of the toroidal and poloidal harmonics need to be stored for each particular configuration \cite{dommaschk1986representations,thomas1997new}, implementing \(\psi_v\) is more difficult. Ideally, \(\psi_v\) should be chosen so that \(\nabla\psi_v\) is continuous. For example, in the case of a tokamak vacuum field \(\vec B_v = F_0\nabla\phi\), one can set \(\psi_v\) to be the minor radius, or, if the vacuum field has flux surfaces, one can set \(\psi_v\) as the flux surface label. In both cases, \(\nabla\psi_v\) will be continuous everywhere, except for the \(\psi_v = 0\) axis. If magnetic islands are present, one can either group them with the induced field, keeping only the component of the vacuum field that has nested flux surfaces in \(\vec B_v\), or \(\psi_v\) can be set to be the poloidal flux. In the latter case, \(\nabla\psi_v\) will be perpendicular to the total vacuum field even in the islands, but it will be discontinuous at the X and O points in addition to the magnetic axis. Finally, in stochastic field regions, both \(\psi_v\) and its gradient will inevitably be discontinuous across the cut. In such a case, we can define \(\psi_v\) as the flux surface label in the central region, where flux surfaces exist. Then, while on the cutting surface, we note the contours formed by the flux surfaces intersecting the cutting surface and extend those contours into the area where the stochastic field intersects the cutting surface. A \(\psi_v\) value will then be assigned to each stochastic field line on the basis of the contour that it intersects the cutting plane through. However, in a finite element code, it may not be possible to represent a discontinuous function in the basis used by that code, and so one may have to approximate \(\psi_v\) with a continuous function, and so \(\psi_v\) will no longer be constant along field lines. However, since the field line diffusion coefficient is small in typical devices \cite{abdullaev2008modelling}, the field line drift will be small compared to machine size, and the error will not be significant. Alternatively, one can forgo computing \(\psi_v\) and just find the components of its gradient by solving the system of linear ordinary differential equations derived by Xanthopoulos and Jenko \cite{xanthopoulos2006clebsch}. However, in this case, since its components will only be known up to numerical accuracy, \(\nabla\psi_v\) will only be a gradient up to numerical accuracy, and so the form of the magnetic field introduced in subsection \ref{sec:rep} will only be divergence-free up to numerical accuracy, not machine precision.

\subsection{Magnetic field and velocity representations}
\label{sec:rep}

For an arbitrary magnetic field, the vector potential can be represented in the vacuum field-aligned Clebsch-type coordinates as
\begin{equation}
\label{eq:vpot}
\vec A = \Psi\nabla\chi + \Omega\nabla\psi_v,
\end{equation}
where the \(\nabla\beta_v\) component was eliminated using a gauge transform. Expressing the vector potential of the induced magnetic field alone in the form \eqref{eq:vpot}, the total magnetic field can be expressed as
\begin{equation}
\label{eq:mfield}
\vec B = \nabla\chi + \nabla\Psi\times\nabla\chi + \nabla\Omega\times\nabla\psi_v.
\end{equation}
Since \(\chi\) satisfies the Laplace equation, this form guarantees that the magnetic field will be divergence free, even when the last term is dropped in the context of reduced MHD. Also note that we have partially fixed the gauge: the gradient of a scalar function \(F\) can only be added to the vector potential if \(\partial F/\partial\beta_v = 0\).

Now consider the velocity field. To separate the three MHD waves, Izzo \emph{et al} \cite{izzo1985reduced} and Breslau \emph{et al} \cite{breslau2009some} used a three-term expansion of the velocity, with each wave contained in a specific term. We generalize their expression to the case of \(\chi \neq F_0\phi\) while keeping the first two terms consistent with \cite{strauss1997reduced,franck2015energy}:
\begin{equation}
\label{eq:vfield}
\vec v = \frac{\nabla\Phi\times\nabla\chi}{B_v^2} + v_\parallel \vec B + \pgrad \zeta,
\end{equation}
where \(\pgrad = \nabla - \llgrad\) and \(\llgrad = \frac{\nabla\chi}{B_v^2}\nabla\chi\cdot\nabla\). The superscripts are used to distinguish the parallel and perpendicular gradients with respect to the vacuum field from those defined with respect to the total field in the equations \eqref{eq:vrmhd}. Our expression matches that of Izzo \emph{et al} and Breslau \emph{et al} in the case of a tokamak vacuum field \(\chi = F_0\phi\), except for the second term, which we have made to match the reduced expressions in \cite{strauss1997reduced,franck2015energy}. Note that unlike the magnetic field in expression \eqref{eq:mfield}, which is in mixed form, the velocity in expression \eqref{eq:vfield} is in contravariant form in the Clebsch-type coordinates aligned to the total magnetic field \(\vec B\), as the cross product in the first term will produce covariant basis vectors, and the gradient in the last term can also be written in terms of covariant basis vectors due to being perpendicular to \(\vec e~^\chi\).

The terms in expressions \eqref{eq:mfield} and \eqref{eq:vfield} can be interpreted as, respectively, the background vacuum field, field line bending and field compression for the magnetic field, and \(\vec E\times\vec B\) velocity, field-aligned flow and fluid compression for the velocity. Note that these interpretations are not exact. In particular, the \(\llgrad\Omega\times\nabla\psi_v\) part of the third term in expression \eqref{eq:mfield} is a correction to the field line bending. However, when \(\vec B_v\) is sufficiently strong, the perpendicular gradients of the hydromagnetic variables will dominate and \(\llgrad\Omega\times\nabla\psi_v\) will be small compared to \(\pgrad\Omega\times\nabla\psi_v\). Also, as will be shown in section \ref{sec:rmhd}, the first term in expression \eqref{eq:vfield} is exactly the \(\vec E\times\vec B\) flow only in ideal MHD. Finally, all terms in expression \eqref{eq:vfield} have nonzero divergence, so the last term is not the exact fluid compression term.

We will now prove that any arbitrary vector field can be expressed in the form \eqref{eq:vfield}. Define three projection operators:
\begin{equation}
\label{eq:projop}
\begin{gathered}
\nabla\chi\cdot\nabla\times[\nabla\chi\times(\vec e_\chi\times \\
\nabla\chi\cdot \\
\nabla\cdot [B_v^2 \nabla\chi\times(\vec e_\chi\times
\end{gathered}
\end{equation}
Note that using the identity \(\nabla a\cdot\nabla\times\vec A = -\nabla\cdot(\nabla a\times\vec A)\), which follows directly from the cross product rule for the divergence, the first projection operator can alternatively be expressed as
\begin{equation}
\label{eq:altpop}
-\nabla\cdot[\nabla\chi\times(\nabla\chi\times(\vec e_\chi\times
\end{equation}
Also, notice that the effect of the \(\nabla\chi\times(\vec e_\chi\times\) sub-operator is to subtract out the contravariant \(\chi\) component of a vector: \(\nabla\chi\times(\vec e_\chi\times\vec s) = -\vec s + s^\chi\vec e_\chi\). Applying each of the three operators to expression \eqref{eq:vfield}, we obtain equations for the scalar functions \(\Phi\), \(v_\parallel\) and \(\zeta\):
\begin{equation}
\label{eq:peqs}
\begin{gathered}
\pLap\Phi = \nabla\chi\cdot\nabla\times[\nabla\chi\times(\vec e_\chi\times\vec v)], \\
v_\parallel = \frac{v^\chi}{B^\chi}, \\
\nabla\cdot(B_v^2 \pgrad \zeta) = -\nabla\cdot [B_v^2\nabla\chi\times(\vec e_\chi\times\vec v)],
\end{gathered}
\end{equation}
where \(\pLap = \nabla\cdot\pgrad\). Thus, for any given \(\vec v\), we have two uncoupled linear differential equations for \(\Phi\) and \(\zeta\) and a direct relation for \(v_\parallel\). Both of the linear differential equations are generalized Poisson equations, and the boundary conditions can be obtained as follows. If \(\vec v\), \(\nabla\chi\) and \(\vec B\) are known everywhere, then a linear combination of Neumann boundary conditions for \(\Phi\) and \(\zeta\) can be determined by subtracting \(v_\parallel\vec B\) from \eqref{eq:vfield}, taking the cross product with \(\nabla\chi\) and then the dot product with the unit normal to the boundary:
\begin{equation}
-\vec n\cdot\pgrad\Phi + \vec n\cdot(\nabla\zeta\times\nabla\chi) = \vec n\cdot[(\vec v - v_\parallel\vec B)\times\nabla\chi], \quad \vec r\in\partial V,
\end{equation}
where \(\vec n\) is the unit normal vector to the boundary and \(\partial V\) is the boundary of the volume \(V\). Having one boundary condition for two equations, we have the freedom to introduce a free function \(f(\vec r)\) as follows:
\begin{equation}
\label{eq:bc1}
\begin{gathered}
\vec n\cdot\pgrad\Phi = f(\vec r), \quad \vec r\in\partial V, \\
\vec n\cdot(\nabla\zeta\times\nabla\chi) = \vec n\cdot[(\vec v - v_\parallel\vec B)\times\nabla\chi] + f(\vec r), \quad \vec r\in\partial V.
\end{gathered}
\end{equation}
The consistency condition for the first equation in \eqref{eq:peqs}
\begin{equation}
\begin{aligned}
\oint\limits_{\partial V} \pgrad\Phi\cdot d\vec S &= \int\limits_V \nabla\chi\cdot\nabla\times[\nabla\chi\times(\vec e_\chi\times\vec v)]dV \\
&= -\oint\limits_{\partial V} [(\vec v - v_\parallel\vec B)\times\nabla\chi]\cdot d\vec S
\end{aligned}
\end{equation}
can be satisfied by requiring \(f(\vec r)\) to satisfy \(\oint_{\partial V} f(\vec r)dS = -\oint_{\partial V} [(\vec v - v_\parallel\vec B)\times\nabla\chi]\cdot d\vec S\). Thus, \(\Phi\) is guaranteed to exist, and, by the uniqueness theorem for Poisson's equation, is unique up to a constant when \(f(\vec r)\) is specified.

For \(\zeta\), another linear combination of Neumann boundary conditions can be obtained by subtracting \(v_\parallel\vec B\) from \eqref{eq:vfield}, multiplying by \(B_v^2\) and taking the dot product with the unit normal to the boundary:
\begin{equation}
\vec n\cdot(\nabla\Phi\times\nabla\chi) + B_v^2\vec n\cdot\pgrad\zeta = B_v^2\vec n\cdot(\vec v - v_\parallel\vec B).
\end{equation}
As before, a free function \(g(\vec r)\) is introduced such that:
\begin{equation}
\label{eq:bc2}
\begin{gathered}
B_v^2\vec n\cdot\pgrad\zeta = g(\vec r), \quad \vec r\in\partial V, \\
\vec n\cdot(\nabla\Phi\times\nabla\chi) = \vec n\cdot(\vec v - v_\parallel\vec B) - g(\vec r), \quad \vec r\in\partial V,
\end{gathered}
\end{equation}
and the consistency condition for the third equation in \eqref{eq:peqs}
\begin{equation}
\begin{aligned}
\oint\limits_{\partial V} B_v^2\pgrad\zeta\cdot d\vec S &= -\int\limits_V \nabla\cdot[B_v^2\nabla\chi\times(\vec e_\chi\times\vec v)]dV \\
&= \oint\limits_{\partial V} B_v^2(\vec v - v_\parallel\vec B)\cdot d\vec S
\end{aligned}
\end{equation}
is satisfied by requiring \(\oint_{\partial V} g(\vec r)dS = \oint_{\partial V} B_v^2(\vec v - v_\parallel\vec B)\cdot d\vec S\). Thus, \(\zeta\) is also well-defined. In hindsight, after the existence of scalar functions \(\Phi\) and \(\zeta\) has been proven, the requirements imposed on \(f(\vec r)\) and \(g(\vec r)\) follow from the second conditions in \eqref{eq:bc1} and \eqref{eq:bc2} by integrating over \(\partial V\) and applying the divergence theorem to the LHS.

\section{MHD waves and velocity representation terms}
\label{sec:waves}

In this section, we show in the framework of linearized MHD that the three MHD waves are approximately separated into three terms by the representation \eqref{eq:vfield}, with each term containing a wave and some instabilities. In addition to the usual assumptions of linearized MHD, we also assume that the induced equilibrium magnetic field is small compared to the vacuum field:
\begin{equation}
\label{eq:assmp}
\frac{|\vec B_0 - \nabla\chi|}{|\nabla\chi|} \ll 1.
\end{equation}
Thus, we can approximate \(\vec B_0\) by \(\nabla\chi\) in the following analysis. At the same time, we do not assume that \(\vec j_0 = \frac{1}{\mu_0}\nabla\times\vec B_0\) is negligible. However, the presence of equilibrium current will make no difference in the forthcoming analysis.

We begin with the linearized ideal MHD equation for velocity:
\begin{equation}
\label{eq:lineqn}
\begin{aligned}
\rho_0\ttderiv{\vec v} &= \vec j_0\times[\nabla\times(\vec v\times\nabla\chi)] \\
&+ \frac{1}{\mu_0}[\nabla\times[\nabla\times(\vec v\times\nabla\chi)]]\times\nabla\chi + \nabla(\vec v\cdot\nabla p_0) \\
&+ \gamma\nabla(p_0\nabla\cdot\vec v).
\end{aligned}
\end{equation}
In general, both fluid-compressional and shear waves can propagate in a plasma, just like in an elastic solid. Comparing equation \eqref{eq:lineqn} with a typical elastic wave equation \cite{hudson1980excitation}, we see that the second term on the RHS has similar structure to the shear wave term in an elastic wave equation (this term can compress the magnetic field, but not the fluid), whereas the last term is similar to the compressional wave term. The other two terms on the RHS of equation \eqref{eq:lineqn} do not have second order derivatives of \(\vec v\).

We now evaluate the effect of each term in equation \eqref{eq:vfield} by direct substitution into equation \eqref{eq:lineqn}. Inserting just the first term from expression \eqref{eq:vfield} into the above equation, one obtains:
\begin{equation}
\label{eq:aeqn}
\begin{aligned}
\frac{\rho_0}{B_v^2}\ttderiv{\nabla\Phi}\times\nabla\chi &= -\vec j_0\times(\nabla\times\pgrad\Phi) \\
&- \frac{1}{\mu_0}[\nabla\times(\nabla\times\pgrad\Phi)]\times\nabla\chi \\
&+ \nabla\left(\frac{1}{B_v}[p_0,\Phi]\right) - 2\gamma\nabla\left(\frac{p_0}{B_v^2}[B_v,\Phi]\right),
\end{aligned}
\end{equation}
where \([a,b] = B_v^{-1}\nabla\chi\cdot(\nabla a\times\nabla b)\) is the Poisson bracket for scalar fields \(a\) and \(b\). The first term of expression \eqref{eq:vfield} allows for shear Alfv\'{e}n waves as well as various instabilities, for example, the first term on the RHS of equation \eqref{eq:aeqn} represents current-driven instabilities. We will only consider waves in this section, as a proper consideration of instabilities is better done with an energy principle. Several terms in equation \eqref{eq:aeqn} also account for distortions of waves due to inhomogeneity in the equilibrium, however we are only interested in identifying the MHD wave contained in each term of expression \eqref{eq:vfield}. Such identification is normally performed by determining the propagation speeds of wave-like perturbations, i.e. by determining the coefficient in the second order derivative (with respect to the unknown) in the wave equations \cite{southwood1985curvature,klimushkin2004toroidal}.

Let \(\pgrad\Phi\), which can be interpreted as the perpendicular component of the electric field in reduced ideal MHD (see section \ref{sec:rmhd}), be the unknown in the vector wave equation. Then the second term on the RHS of \eqref{eq:aeqn}, which is the only term that contains second order derivatives of \(\pgrad\Phi\), describes the propagation of waves. Note that what was the fluid-compressional wave term in equation \eqref{eq:lineqn} no longer contains second order derivatives of \(\pgrad\Phi\) in equation \eqref{eq:aeqn}. We focus attention on the second term, which can be rewritten as
\begin{equation}
-\frac{1}{\mu_0}[\pgrad(\pLap\Phi)-\pLap(\pgrad\Phi)-\llLap(\pgrad\Phi)]\times\nabla\chi.
\end{equation}
We used the fact that \(\nabla\times(\nabla\times\vec A) = \nabla(\nabla\cdot\vec A) - \Delta\vec A\), \(\nabla = \pgrad + \llgrad\) and \(\Delta = \pLap + \llLap\), where \(\llLap = \nabla\cdot\llgrad\). Now consider the two terms \(\pgrad(\pLap\Phi)\) and \(-\pLap(\pgrad\Phi)\). The operators \(\pgrad\) and \(\pLap\) do not commute when the background field is not uniform, however, as can be shown by lengthy algebraic manipulations, the highest order derivatives do indeed cancel, leaving only first and second order derivatives of \(\Phi\). Thus, the only term containing third order derivatives of \(\Phi\) (i.e. second order derivatives of \(\pgrad\Phi\)) is \(-\llLap(\pgrad\Phi)\). Cross multiplying equation \eqref{eq:aeqn} by \(\nabla\chi\) from the the left and dividing by \(\rho_0\), one obtains
\begin{equation}
\begin{aligned}
\ttderiv{\pgrad\Phi} &= \frac{B_v^2}{\mu_0\rho_0}\left[\llLap(\pgrad\Phi) + [\pLap,\pgrad]\Phi\right] \\
&- \frac{\nabla\chi}{\mu_0\rho_0}\nabla\chi\cdot\left[[\llLap,\pgrad]\Phi + [\pLap,\pgrad]\Phi\right] \\
&+ \frac{j_0^\chi}{\rho_0}\nabla\times\pgrad\Phi + \frac{1}{\rho_0}\nabla\chi\times\nabla\left(\frac{1}{B_v}[p_0,\Phi]\right) \\
&- \frac{2\gamma}{\rho_0}\nabla\chi\times\nabla\left(\frac{p_0}{B_v^2}[B_v,\Phi]\right),
\end{aligned}
\end{equation}
where, by a slight abuse of notation, square brackets applied to two operators \(A, B\) that act on a function \(f\) is understood as the commutator of the operators: \([A,B]f = A(Bf) - B(Af)\). Similarly to \([\pLap,\pgrad]\), \([\llLap,\pgrad]\) does not contain third order derivatives. Clearly, waves in the vector field \(\pgrad\Phi\) will propagate along field lines with the Alfv\'{e}n speed, while the velocity perturbation is perpendicular to the field lines. Thus, shear Alfv\'{e}n waves are the only MHD waves allowed by the first term of expression \eqref{eq:vfield}. Note that, due to magnetic field nonuniformity, the first term in expression \eqref{eq:vfield} has nonzero divergence. Although the Alfv\'{e}n wave in a uniform background field is incompressible, this is not the case in a curved field \cite{southwood1985curvature}.

We now consider the second term in the expression \eqref{eq:vfield}. Note that in the context of linearized MHD with assumption \eqref{eq:assmp} the second term can be approximated as:
\begin{equation}
\label{eq:aprx}
v_\parallel\vec B = v_\parallel (\vec B_0 + \vec B_1) \approx v_\parallel\nabla\chi,
\end{equation}
since both velocity and \(\vec B_1\) are first-order quantities. However, while using the full field instead of just the vacuum field in the second term makes no difference from the linear MHD wave point of view, the full field provides the advantage of allowing temperature and density profiles to flatten when the field becomes stochastic in a certain area, which would not be so simple in reduced MHD (see section \ref{sec:rmhd}) if the parallel velocity was directed along the static background field.

Inserting the approximation for the second term \eqref{eq:aprx} into equation \eqref{eq:lineqn}, one obtains
\begin{equation}
\rho_0\ttderiv{v_\parallel}\nabla\chi = \gamma\nabla (B_v p_0\llderiv v_\parallel),
\end{equation}
where \(\llderiv = B_v^{-1}\nabla\chi\cdot\nabla\) is the spatial derivative along the vacuum field. The third term on the RHS of equation \eqref{eq:lineqn} was dropped due to the fact that \(\llderiv p_0 \approx 0\) since \(\nabla p_0 = \vec j_0\times\vec B_0\) and \(\vec B_0 \approx \nabla\chi\). Note that only the fluid-compressional term from equation \eqref{eq:lineqn} survives. Expanding the RHS, multiplying by \(\nabla\chi\) and dividing by \(\rho_0 B_v^2\), one obtains
\begin{equation}
\begin{aligned}
\ttderiv{v_\parallel} &= \frac{\gamma p_0}{\rho_0}(\llderiv)^2 v_\parallel + \frac{\gamma p_0}{\rho_0 B_v}\llderiv v_\parallel \llderiv B_v \\
&= \frac{\gamma p_0}{\rho_0}\llLap v_\parallel + \frac{2\gamma p_0}{\rho_0 B_v}\llderiv v_\parallel \llderiv B_v.
\end{aligned}
\end{equation}
We see that waves in the scalar field \(v_\parallel\) propagate with the sound speed along field lines while the velocity perturbation is parallel to the field lines. Thus, only slow magnetosonic waves are allowed by the second term of expression \eqref{eq:vfield}. The reason why these waves propagate with the sound speed instead of the slow magnetosonic speed is because we have constrained the velocity perturbation to be parallel to the background field, zeroing out the shear term and making it impossible for the wave to compress the magnetic field. Being able to only compress the fluid, the wave behaves as a sound wave. A true slow magnetosonic wave can exist when the third term of the expression \eqref{eq:vfield} is also included, due to coupling between the second and third terms.

We now show that the first two terms in expression \eqref{eq:vfield} do not compress the magnetic field even in the nonlinear regime. We start with the ideal MHD induction equation, insert expressions \eqref{eq:mfield} and the first two terms of \eqref{eq:vfield}. Multiplying by \(\nabla\chi\) gives the component of \(\partial\vec B/\partial t\) along the vacuum magnetic field, which corresponds to field compression:
\begin{equation}
\begin{aligned}
\tderiv{B^\chi} &= B_v\left[\tderiv{\Omega},\psi_v\right] = \nabla\chi\cdot\nabla\times(\vec v\times\vec B) \\
&= \nabla\cdot\Bigg[\Bigg(-\nabla\Phi + \frac{\nabla\chi}{B_v^2}\nabla\chi\cdot\nabla\Phi - \frac{\nabla\chi}{B_v}[\Psi,\Phi] \\
&+ \frac{\nabla\Omega}{B_v}[\psi_v,\Phi] - \frac{\nabla\psi_v}{B_v}[\Omega,\Phi]\Bigg)\times\nabla\chi\Bigg] \\
&= B_v\left[\frac{[\psi_v,\Phi]}{B_v},\Omega\right] - B_v\left[\frac{[\Omega,\Phi]}{B_v},\psi_v\right].
\end{aligned}
\end{equation}
As was mentioned in subsection \ref{sec:rep}, \(\pgrad\Omega\) corresponds to field compression, and \(\llgrad\Omega\) will not contribute to the Poisson brackets above. From the above equation, if \(\pgrad\Omega = 0\) initially, then \(\pgrad\partial\Omega/\partial t = 0\), and so \(\pgrad\Omega\) will stay at zero. Thus, if there is no compression initially, the first two terms of expression \eqref{eq:vfield} will not produce any compression. This is yet another advantage of using the full field in the second term of expression \eqref{eq:vfield} instead of just the vacuum field. We note here that describing equilibria with a Shafranov shift is impossible without including vacuum field compression, and so setting \(\Omega = 0\) eliminates the possibility of consistently using tokamak equilibria obtained from the Grad-Shafranov equation as initial conditions. It also makes it impossible to directly use most stellarator equilibria, as those also include a Shafranov shift. One possibility is to obtain the initial conditions simply by neglecting the equilibrium field compression, as done in JOREK. In the case of a tokamak, this amounts to solving the Grad-Shafranov equation and using the results as initial conditions for \(\Psi\) and \(p\), but keeping \(\Omega = 0\), which is incompatible with the Grad-Shafranov expression for the magnetic field. Another possibility would be to allow field compression by evolving \(\Omega\) at the expense of more unknowns and more complicated equations, as done in the model by Izzo \emph{et al} and the M3D-$C^1$ four-field model \cite{izzo1985reduced,breslau2009some}. However, it is usually advantageous to use the first method, as neglecting equilibrium field compression, which is anyway small due to \eqref{eq:assmp}, guarantees that the field will not be further compressed during the simulation, as shown above. Since field compression requires significant energy, there is a large restoring force associated with it, and unstable modes will generally not compress the field \cite{ferraro2011fluid}. Thus, it makes sense to consistently neglect \(\Omega\).

Finally, we consider the last term in the expression \eqref{eq:vfield}. Inserting it into equation \eqref{eq:lineqn}, we obtain:
\begin{equation}
\label{eq:lterm}
\begin{aligned}
\rho_0\ttderiv{\pgrad\zeta} &= \vec j_0\times[\nabla\times(\nabla\zeta\times\nabla\chi)] \\
&- \frac{1}{\mu_0}\Delta(\nabla\zeta\times\nabla\chi)\times\nabla\chi + \nabla(\zeta, p_0) \\
&+ \gamma\nabla(p_0\pLap\zeta),
\end{aligned}
\end{equation}
where \((a,b) = \pgrad a\cdot\pgrad b\) is the inner product of the perpendicular gradients of \(a\) and \(b\). We take \(\pgrad\zeta\) to be the unknown in the wave equation, the which is given by the perpendicular component of equation \eqref{eq:lterm}. Taking the perpendicular component of the equation, dividing by \(\rho_0\) and using \(\Delta(\vec A\times\vec B) = (\Delta\vec A)\times\vec B + 2\nabla\vec A\dotcross\nabla\vec B + \vec A\times(\Delta\vec B)\) to expand the RHS, we get:
\begin{equation}
\label{eq:fwave}
\begin{aligned}
\ttderiv{\pgrad\zeta} &= \frac{\vec j_0}{\rho_0}\times[\nabla\times(\nabla\zeta\times\nabla\chi)] \\
&- \frac{\nabla\chi}{B_v\rho_0}\cdot[\vec j_0\times[\nabla\times(\nabla\zeta\times\nabla\chi)]] \\
&+ \frac{B_v^2}{\mu_0\rho_0}\Delta(\pgrad\zeta) - \frac{\nabla\chi}{\mu_0\rho_0}(\nabla\chi\cdot[\Delta,\pgrad]\zeta) \\
&+ \frac{2}{\mu_0\rho_0}\nabla\chi\times(\nabla\pgrad\zeta\dotcross\nabla\nabla\chi)  + \frac{1}{\rho_0}\pgrad(\zeta,p_0) \\
&+ \frac{\gamma}{\rho_0}\pLap\zeta\pgrad p_0 + \frac{\gamma p_0}{\rho_0}\pLap(\pgrad\zeta) \\
&+ \frac{\gamma p_0}{\rho_0}[\pgrad,\pLap]\zeta,
\end{aligned}
\end{equation}
where \(\overleftrightarrow{T}\dotcross\overleftrightarrow{U} = \epsilon_{ijk}JT^{lj}U_l~^k\vec e~^i\) is the dot-cross product of two tensors, \(\epsilon_{ijk}\) is the Levi-Civita symbol and \(J\) is the Jacobian. Just as before, the commutators do not contain third order derivatives. Only the third and second to last terms on the RHS of the above equation contain second order derivatives of \(\pgrad\zeta\) (third order derivatives of \(\zeta\)). The propagation of waves is described by these two terms, which can be rewritten as
\begin{equation}
\left(\frac{B_v^2}{\mu_0\rho_0} + \frac{\gamma p_0}{\rho_0}\right)\pLap(\pgrad\zeta) + \frac{B_v^2}{\mu_0\rho_0}\llLap(\pgrad\zeta).
\end{equation}
The wave can propagate both along and across field lines with different speeds. Now consider the fast magnetosonic wave speed in the form presented by Freidberg \cite{freidberg2014ideal}:
\begin{equation}
\label{eq:fwspeed}
c_f = \sqrt{\frac{1}{2}(c_A^2 + c_s^2)\left(1 + \sqrt{1 - 4\cos^2\theta\frac{c_A^2 c_s^2}{(c_A^2+c_s^2)^2}}\right)},
\end{equation}
where \(\theta\) is the angle between the direction of wave propagation and the field, \(c_A = B_0/\sqrt{\mu_0\rho_0}\) is the Alfv\'{e}n speed and \(c_s = \sqrt{\gamma p_0/\rho_0}\) is the sound speed. To transform Freidberg's original expression, which was written in terms of the wave number \(k\), the fact that \(k_\parallel = k\cos\theta\) was used. We see that the speed of the wave in equation \eqref{eq:fwave} matches \eqref{eq:fwspeed} both in the parallel (\(c_w=c_f=c_A\)) and perpendicular (\(c_w=c_f=\sqrt{c_A^2+c_s^2}\)) directions, given that \(B_0 \approx B_v\), as implied by assumption \eqref{eq:assmp}. In general, however, the speed of the wave in equation \eqref{eq:fwave} is
\begin{equation}
\label{eq:fwest}
\begin{aligned}
c_w &= \sqrt{(c_A^2 + c_s^2)\sin^2\theta + c_A^2\cos^2\theta} \\
&= \sqrt{c_A^2 + c_s^2 - c_s^2\cos^2\theta}.
\end{aligned}
\end{equation}
For \(\beta < 1\), we have \(c_s < c_A\), and as \(\beta \to 0\) both \(c_f \to c_A\) and \(c_w \to c_A\). In the \(c_s \leq c_A\) regime, the discrepancy in the fast magnetosonic wave speed estimated by \eqref{eq:fwest} will be maximized to about 9\% when \(c_s = c_A\), which corresponds to \(\beta = 2/\gamma\), and when \(\theta = \arccos(\pm\sqrt{2\sqrt{2}-2})\). The discrepancy arises because the third term of expression \eqref{eq:vfield} was constrained to be orthogonal to magnetic field. Just like the slow magnetosonic wave, a true fast magnetosonic wave will only occur via coupling between the second and third terms. Nevertheless, the approximate separation of the slow and fast waves provided by second and third terms is fairly accurate for low \(\beta\), and the third term does manage to separate out the fast \(\sqrt{c_A^2+c_s^2}\) dynamics, the removal of which decreases the stiffness of the equations and is sufficient for the purposes of reduced MHD (see section \ref{sec:rmhd}).

\section{Derivation of the equations}
\label{sec:derivation}

In this section, we derive scalar equations for the potentials \(\Psi\), \(\Omega\), \(\Phi\) and \(\zeta\), and the parallel component \(v_\parallel\) from the vector equations in \eqref{eq:vrmhd} by inserting expressions \eqref{eq:mfield} and \eqref{eq:vfield} into them and applying projection operators. Since any arbitrary magnetic field and velocity can be represented in the forms \eqref{eq:mfield} and \eqref{eq:vfield}, the scalar equations that we derive are still full MHD equations. The reduction procedure is applied separately from the derivation in section \ref{sec:rmhd}.

The continuity and energy conservation equations \eqref{eq:vrmhd}, which we will use to evolve  density and pressure, can be employed directly by inserting expressions \eqref{eq:mfield} and \eqref{eq:vfield} into them:
\begin{widetext}
\begin{equation}
\label{eq:rhoeq}
\tderiv{\rho} = - B_v\left[\frac{\rho}{B_v^2},\Phi\right] - B_v\llderiv(\rho v_\parallel) - B_v[\rho v_\parallel,\Psi] - F_v[\rho v_\parallel,\Omega]_{\psi_v} - (\rho,\zeta) - \rho\pLap\zeta + P,
\end{equation}
and
\begin{equation}
\label{eq:eneq}
\begin{aligned}
&\tderiv{}\left(\frac{\rho v^2}{2} + \frac{p}{\gamma-1} + \frac{B^2}{2\mu_0}\right) = - B_v\left[\frac{1}{B_v^2}\left(\frac{\rho v^2}{2} + \frac{\gamma p}{\gamma-1} + \frac{B^2}{\mu_0}\right),\Phi\right] - B_v\llderiv\left[v_\parallel\left(\frac{\rho v^2}{2} + \frac{\gamma p}{\gamma-1} + \frac{B^2}{\mu_0}\right) - \frac{\vec v\cdot\vec B}{\mu_0}\right] \\
&- B_v\left[v_\parallel\left(\frac{\rho v^2}{2} + \frac{\gamma p}{\gamma-1} + \frac{B^2}{\mu_0}\right) - \frac{\vec v\cdot\vec B}{\mu_0},\Psi\right] - F_v\left[v_\parallel\left(\frac{\rho v^2}{2} + \frac{\gamma p}{\gamma-1} + \frac{B^2}{\mu_0}\right) - \frac{\vec v\cdot\vec B}{\mu_0},\Omega\right]_{\psi_v} \\
&- \left(\frac{\rho v^2}{2} + \frac{\gamma p}{\gamma-1} + \frac{B^2}{\mu_0},\zeta\right) - \left(\frac{\rho v^2}{2} + \frac{\gamma p}{\gamma-1} + \frac{B^2}{\mu_0}\right)\pLap\zeta - \frac{1}{\mu_0}\nabla\cdot(\eta\vec j\times\vec B) \\
&+ \nabla\cdot\Bigg[\frac{p}{\gamma-1}\frac{D_\perp}{\rho}\left(\nabla\rho-\frac{\vec B}{B^2}(B_v\llderiv\rho + B_v[\rho,\Psi] + F_v[\rho,\Omega]_{\psi_v})\right) \\
&+ \frac{\kappa_\perp}{R}\nabla\left(\frac{p}{\rho}\right) + \frac{(\kappa_\parallel-\kappa_\perp)\vec B}{RB^2}\left(B_v\llderiv\left(\frac{p}{\rho}\right) + B_v\left[\frac{p}{\rho},\Psi\right] + F_v\left[\frac{p}{\rho},\Omega\right]_{\psi_v}\right)\Bigg] + S_e - \frac{v^2}{2}P,
\end{aligned}
\end{equation}
where \(F_v = |\nabla\psi_v|\) and \([a,b]_{\psi_v} = F_v^{-1}\nabla\psi_v\cdot(\nabla a\times\nabla b)\) is the Poisson bracket for scalar fields \(a\) and \(b\) with respect to \(\nabla\psi_v\). In addition,
\begin{equation}
\label{eq:exprs}
\begin{gathered}
P = \nabla\cdot\left[D_\perp\nabla\rho - \frac{D_\perp\vec B}{B^2}(B_v\llderiv\rho + B_v[\rho,\Psi] + F_v[\rho,\Omega]_{\psi_v})\right] + S_\rho, \\
v^2 = \frac{1}{B_v^2}(\Phi,\Phi) + 2v_\parallel\vec v\cdot\vec B + \frac{2}{B_v}[\zeta,\Phi] - v_\parallel^2 B^2 + (\zeta,\zeta), \\
\vec v\cdot\vec B = (\Phi,\Psi) - \frac{F_v}{B_v}\psderiv\Phi\llderiv\Omega + v_\parallel B^2 + B_v[\zeta,\Psi] + F_v[\zeta,\Omega]_{\psi_v} - \llderiv\zeta[\Omega,\psi_v], \\
B^2 = B_v^2 + 2B_v[\Omega,\psi_v] + B_v^2(\Psi,\Psi) - 2B_vF_v\psderiv\Psi\llderiv\Omega + F_v^2(\Omega,\Omega)_{\psi_v}, \\
\vec j = \frac{1}{\mu_0}[-\nabla\chi\Delta\Psi + B_v\llderiv\nabla\Psi - (\nabla\Psi\cdot\nabla)\nabla\chi + \nabla\Omega\Delta\psi_v - \nabla\psi_v\Delta\Omega + F_v\psderiv\nabla\Omega - (\nabla\Omega\cdot\nabla)\nabla\psi_v].
\end{gathered}
\end{equation}
where \(\psderiv = F_v^{-1}\nabla\psi_v\cdot\nabla\) is the spatial derivative in the direction of \(\nabla\psi_v\), and \((a,b)_{\psi_v} = \nabla a\cdot\nabla b - \psderiv a\psderiv b\) is the inner product of gradients of scalar functions \(a\) and \(b\) perpendicular to \(\nabla\psi_v\).

We now proceed to derive the scalar equations for the magnetic potentials. Inserting expressions \eqref{eq:mfield} and \eqref{eq:vfield} into the induction equation \eqref{eq:vrmhd}, we obtain:
\begin{equation}
\label{eq:vmfeq}
\begin{aligned}
\nabla\tderiv{\Psi}\times\nabla\chi + \nabla\tderiv{\Omega}\times\nabla\psi_v &= \nabla\times\Bigg[\frac{\nabla\chi}{B_v}(\llderiv\Phi - [\Psi,\Phi]) + \frac{\nabla\Omega}{B_v}[\psi_v,\Phi] - \frac{\nabla\psi_v}{B_v}[\Omega,\Phi] + \nabla\zeta\times\nabla\chi \\
&- \nabla\chi(\zeta,\Psi) + \nabla\Omega(\zeta,\psi_v) - \nabla\psi_v(\zeta,\Omega) - \eta\vec j\Bigg].
\end{aligned}
\end{equation}
Projecting this vector equation on the \(\nabla\psi_v\) and the \(\nabla\chi\) directions, we obtain scalar evolution equations for the \(\Psi\) and \(\Omega\) potentials:
\begin{equation}
\label{eq:mfeqns}
\begin{aligned}
\left[\psi_v,\tderiv{\Psi}\right] &= \left[\frac{[\Psi,\Phi]-\llderiv\Phi}{B_v},\psi_v\right] - \frac{F_v}{B_v}\left[\Omega,\frac{[\psi_v,\Phi]}{B_v}\right]_{\psi_v} + \llderiv(F_v\psderiv\zeta) + [(\zeta,\Psi),\psi_v] - \frac{F_v}{B_v}[\Omega,(\zeta,\psi_v)]_{\psi_v} \\
&+ \frac{1}{B_v}\nabla\cdot(\eta\nabla\psi_v\times\vec j), \\
\left[\tderiv{\Omega},\psi_v\right] &= -\left[\Omega,\frac{[\psi_v,\Phi]}{B_v}\right] + \left[\psi_v,\frac{[\Omega,\Phi]}{B_v}\right] - 2(B_v,\zeta) - B_v\pLap\zeta - [\Omega,(\zeta,\psi_v)] + [\psi_v,(\zeta,\Omega)] \\
&+ \frac{1}{B_v}\nabla\cdot(\eta\nabla\chi\times\vec j).
\end{aligned}
\end{equation}
\end{widetext}
where we have used the same identity as in \eqref{eq:altpop} to simplify the projections on the right hand side. We could also have obtained evolution equations for \(\Psi\) and \(\Phi\) by using the potential form of Faraday's law, however in that case, due to our choice for the magnetic vector potential \eqref{eq:vpot}, we are no longer free in our choice of the electric potential \(V\), which must be chosen so that \(\partial V/\partial\beta_v\) cancels with \(E_{\beta_v}\), the covariant \(\beta_v\) component of the electric field. This would produce complicated integro-differential equations for \(\Psi\) and \(\Omega\).

To obtain the scalar equations for the potentials \(\Phi\) and \(\zeta\) and the parallel component \(v_\parallel\), we begin by inserting expressions \eqref{eq:mfield} and \eqref{eq:vfield} into the Navier-Stokes equation \eqref{eq:vrmhd}, dividing by \(\rho\) and then apply the projection operators \eqref{eq:projop}. Expanding the time derivative, dividing by \(\rho\) and inserting the appropriate expressions, we have:
\begin{equation}
\label{eq:nsins}
\begin{aligned}
&\nabla\tderiv{\Phi}\times\frac{\nabla\chi}{B_v^2} + \tderiv{v_\parallel}\vec B + v_\parallel\tderiv{\vec B} + \pgrad\tderiv{\zeta} + \frac{\vec v}{\rho}P \\
&+ \frac{1}{2}\nabla v^2 + \vec\omega\times\vec v = \frac{1}{\rho}\vec j\times\vec B - \frac{1}{\rho}\nabla p,
\end{aligned}
\end{equation}
where we used the identity \((\vec v\cdot\nabla)\vec v = \frac{1}{2}\nabla v^2 + \vec\omega\times\vec v\) and \(\vec\omega\) is the vorticity:
\begin{equation}
\begin{aligned}
\vec\omega &= \nabla\times\vec v = -\nabla\chi\nabla\cdot\left(\frac{\nabla\Phi}{B_v^2}\right) + B_v\llderiv\frac{\nabla\Phi}{B_v^2} \\
&- \frac{1}{B_v^2}(\nabla\Phi\cdot\nabla)\nabla\chi + \nabla v_\parallel\times\vec B + v_\parallel\vec j + \nabla\chi\times\nabla\frac{\llderiv\zeta}{B_v}.
\end{aligned}
\end{equation}
As discussed in subsection \ref{sec:vrmhd}, the viscous term is not treated in this derivation. Applying the \(\nabla\chi\times(\vec e_\chi\times\) sub-operator to equation \eqref{eq:nsins}, we obtain:
\begin{widetext}
\begin{equation}
\label{eq:pcsub}
\begin{aligned}
&-\nabla\tderiv{\Phi}\times\frac{\nabla\chi}{B_v^2} - v_\parallel\nabla\tderiv{\Psi}\times\nabla\chi - v_\parallel\nabla\tderiv{\Omega}\times\nabla\psi_v + \vec e_\chi v_\parallel B_v\left[\tderiv{\Omega},\psi_v\right] - \pgrad\tderiv{\zeta} = \frac{1}{2}\nabla v^2 - \vec e_\chi\frac{B_v}{2}\llderiv v^2 \\
&+ \left(\frac{\nabla\Phi\times\nabla\chi}{B_v^2} + \pgrad\zeta\right)\frac{P}{\rho} - v_\chi\nabla\chi\times\vec\omega + \omega_\chi\nabla\chi\times\vec v + \frac{B^2}{\rho B^\chi}\nabla\chi\times\vec j - \frac{j_\chi}{\rho}\nabla\chi\times\vec B + \frac{\nabla p - \vec e_\chi B_v\llderiv p}{\rho}.
\end{aligned}
\end{equation}
Proceeding to obtain the equation for \(\Phi\), we apply the remainder of the projection operator for \(\Phi\), namely \(\nabla\chi\cdot(\nabla\times\), or its equivalent \(-\nabla\cdot(\nabla\chi\times\), when appropriate, to equation \eqref{eq:pcsub}:
\begin{equation}
\label{eq:phieq}
\begin{aligned}
&\pLap\tderiv{\Phi} + \left(v_\parallel B_v^2,\tderiv{\Psi}\right) + v_\parallel B_v^2\pLap\tderiv{\Psi} - F_v\psderiv\left(v_\parallel B_v\llderiv\tderiv{\Omega}\right) - v_\parallel B_v\llderiv\tderiv{\Omega}\Delta\psi_v - \left(\frac{v_\parallel B_v^2}{B_v + [\Omega,\psi_v]}\left[\tderiv{\Omega},\psi_v\right],\Psi\right) \\
&- \frac{v_\parallel B_v^2}{B_v + [\Omega,\psi_v]}\left[\tderiv{\Omega},\psi_v\right]\pLap\Psi + F_v\psderiv\left(\frac{v_\parallel B_v\llderiv\Omega}{B_v + [\Omega,\psi_v]}\left[\tderiv{\Omega},\psi_v\right]\right) + \frac{v_\parallel B_v\llderiv\Omega}{B_v + [\Omega,\psi_v]}\left[\tderiv{\Omega},\psi_v\right]\Delta\psi_v \\
&= \nabla\cdot\Bigg[\frac{B_v\pgrad\Psi - \llderiv\Omega\nabla\psi_v}{B_v + [\Omega,\psi_v]}\frac{B_v\llderiv v^2}{2} + \omega_\chi(\nabla\Phi\times\nabla\chi + B_v^2 v_\parallel\vec B^\perp + B_v^2\pgrad\zeta) - v_\chi B_v^2\vec\omega^\perp + \frac{B^2 B_v}{\rho(B_v+[\Omega,\psi_v])}\vec j^\perp \\
&- \frac{j_\chi B_v^2}{\rho}\vec B^\perp - (\pgrad\Phi + \nabla\chi\times\nabla\zeta)\frac{P}{\rho}\Bigg] + B_v\left[\frac{1}{\rho},p\right] + \left(\frac{B_v^2\llderiv p}{\rho(B_v+[\Omega,\psi_v])},\Psi\right) + \frac{B_v^2\llderiv p}{\rho(B_v+[\Omega,\psi_v])}\pLap\Psi \\
&- F_v\psderiv\left(\frac{B_v\llderiv\Omega\llderiv p}{\rho(B_v+[\Omega,\psi_v])}\right) - \frac{B_v\llderiv\Omega\llderiv p}{\rho(B_v+[\Omega,\psi_v])}\Delta\psi_v + \nu\Delta\pLap\Phi,
\end{aligned}
\end{equation}
\end{widetext}
where we have added a generic viscosity term \(\nu\Delta\pLap\Phi\), as discussed in subsection \ref{sec:vrmhd}. Following the same approach that was used by Franck \emph{et al} \cite{franck2015energy}, we choose the form of the viscosity term by allowing the projection operator to act directly on \(\vec v\), bypassing the \(\nu\Delta\) operator. Here, for any vector \(\vec s\), \(\vec s^\perp = \vec s - s^\chi\nabla\chi/B_v^2\) is the vector component of \(\vec s\) perpendicular to the vacuum field. Subscripts and superscripts on \(v, \omega, j\) and \(B\) denote covariant and contravariant components, respectively.

To get the equation for the parallel component \(v_\parallel\), we apply the projection operator \(\nabla\chi\cdot\) to equation \eqref{eq:nsins} and divide by \(B^\chi\):
\begin{widetext}
\begin{equation}
\label{eq:vpeq}
\begin{aligned}
&\tderiv{v_\parallel} + \frac{v_\parallel}{B_v + [\Omega,\psi_v]}\left[\tderiv{\Omega},\psi_v\right] = -\frac{v_\parallel}{\rho}P + \frac{1}{B_v + [\Omega,\psi_v]}\Bigg[\frac{\vec\omega\cdot\pgrad\Phi}{B_v} + B_v v_\parallel\vec\omega\cdot\pgrad\Psi - v_\parallel\omega^{\psi_v}\llderiv\Omega - \frac{\vec\omega\cdot(\nabla\zeta\times\nabla\chi)}{B_v} \\
&- \frac{\llderiv v^2}{2} - \frac{B_v\vec j\cdot\pgrad\Psi + j^{\psi_v}\llderiv\Omega - \llderiv p}{\rho}\Bigg] + \nu\Delta v_\parallel,
\end{aligned}
\end{equation}
where we have again allowed the projection operator \((B^\chi)^{-1}\nabla\chi\cdot\) to act directly on \(\vec v\).

Finally, to get the equation for \(\zeta\), we apply the remainder of the projection operator for \(\zeta\), namely \(\nabla\cdot(B_v^2\) to equation \eqref{eq:pcsub}:
\begin{equation}
\label{eq:zetaeq}
\begin{aligned}
&B_v^2\pLap\tderiv{\zeta} + \left(B_v^2,\tderiv{\zeta}\right) + B_v\left[v_\parallel B_v^2,\tderiv{\Psi}\right] + F_v\left[v_\parallel B_v^2,\tderiv{\Omega}\right]_{\psi_v} - B_v\llderiv\left(\frac{v_\parallel B_v^2}{B_v + [\Omega,\psi_v]}\left[\tderiv{\Omega},\psi_v\right]\right) \\
&- B_v\left[\frac{v_\parallel B_v^2}{B_v + [\Omega,\psi_v]}\left[\tderiv{\Omega},\psi_v\right],\Psi\right] - F_v\left[\frac{v_\parallel B_v^2}{B_v + [\Omega,\psi_v]}\left[\tderiv{\Omega},\psi_v\right],\Omega\right]_{\psi_v} = \nabla\cdot\Bigg[B_v^2 v_\chi\nabla\chi\times\vec\omega - \frac{B_v^2}{2}\nabla v^2 \\
&-B_v^2\omega_\chi(\pgrad\Phi + v_\parallel B_v^2\pgrad\Psi - v_\parallel B_v\llderiv\Omega\nabla\psi_v + \nabla\chi\times\nabla\zeta) - (\nabla\Phi\times\nabla\chi + B_v^2\pgrad\zeta)\frac{P}{\rho} - \frac{B_v B^2}{\rho(B_v+[\Omega,\psi_v])}\nabla\chi\times\vec j \\
&+ \frac{B_v^3 j_\chi}{\rho}(B_v\pgrad\Psi - \llderiv\Omega\nabla\psi_v) + \frac{B_v^2(\nabla\chi+\nabla\Psi\times\nabla\chi+\nabla\Omega\times\nabla\psi_v)\llderiv v^2}{2(B_v+[\Omega,\psi_v])}\Bigg] - \frac{B_v^2}{\rho}\Delta p - \nabla\left(\frac{B_v^2}{\rho}\right)\cdot\nabla p \\
&+ B_v\llderiv\left(\frac{B_v^2\llderiv p}{\rho(B_v + [\Omega,\psi_v])}\right) + B_v\left[\frac{B_v^2\llderiv p}{\rho(B_v + [\Omega,\psi_v])},\Psi\right] + F_v\left[\frac{B_v^2\llderiv p}{\rho(B_v + [\Omega,\psi_v])},\Omega\right]_{\psi_v} + \nu B_v^2\Delta\pLap\zeta.
\end{aligned}
\end{equation}
\end{widetext}
In this equation, in addition to allowing the projection operator to act directly on \(\vec v\), we also allow \(B_v^2\) to pass through the divergence and Laplacian operators after the projection operator has acted on \(\vec v\). This does not introduce any new error since we already allowed \(B_v^2\) to pass through the Laplacian as part of the projection operator. Strictly speaking, for non-negligible viscosity, the approximations applied to the viscosity term in equations \eqref{eq:phieq}, \eqref{eq:vpeq} and \eqref{eq:zetaeq} are only valid when both the vacuum field and the full magnetic field are approximately uniform, however the viscosity term in the Navier-Stokes equation \eqref{eq:vrmhd} does not accurately model viscous effects in a plasma anyway, and the approximations should still give the correct order of magnitude even when the magnetic field cannot be approximated as uniform \cite{franck2015energy}.

We point out that third- and fourth-order spatial derivatives arise in equations \eqref{eq:mfeqns}, \eqref{eq:phieq} and \eqref{eq:zetaeq}. This can be problematic when the unknown functions are interpolated with third-order polynomials, an approach used by the JOREK code \cite{czarny2008bezier}. To mitigate this problem, we express terms with third order derivatives as divergences and terms with fourth order derivatives as Laplacians, which allows one to reduce the order of the derivatives by applying integration by parts in the weak form of the equations. Finally, we note that, except for the viscosity term, no other approximations were made in this section, and the equations we derived still correspond to full MHD, albeit in a potential form.

Finally, from the form of equation \eqref{eq:eneq}, it may seem that at \(\beta \ll 1\) the magnetic energy is the dominant term in the time derivative. Since equation \eqref{eq:eneq} is used to evolve the pressure, which requires subtracting the changes in kinetic and magnetic energy from the change in total local energy over one time step, this would lead to a large numerical error in the pressure change due to being mixed with the error in the magnetic energy change. Fortunately, this is not the case. If we apply an ordering, then the time derivatives of the kinetic, internal and magnetic energies will all have exactly the same order in \(\lambda\), as shown in appendix \ref{sec:ordering}. Without a strict ordering, we can still argue that the time derivatives of internal and magnetic energy will be comparable. Using the expression for \(B^2\) given in \eqref{eq:exprs} we have:
\begin{equation*}
\frac{\partial}{\partial t}\left(\frac{B^2}{2\mu_0}\right) \sim \frac{B_v^2}{2\mu_0}\frac{\partial}{\partial t}(\Psi,\Psi),
\end{equation*}
where we have assumed that \(\partial\Omega/\partial t\) is negligible. This is because field compression is associated with a large restoring force. We have also assumed that the magnitude of the \(\nabla\Omega\times\nabla\psi_v\) term in the magnetic field is of the same order or smaller than the magnitude of the \(\nabla\Psi\times\nabla\chi\) term. We can estimate the relative magnitudes of \(p\) and \(\Psi\) from the Grad-Shafranov equation (the orders of magnitude should be about the same for a stellarator):
\begin{equation*}
\begin{gathered}
\Delta^*\Psi = -\frac{\mu_0 R^2}{F_0^2}\frac{dp}{d\Psi} - \frac{F}{F_0}\frac{dF}{d\Psi} \\
\Rightarrow p \sim \frac{F_0^2}{\mu_0 R^2}\Psi\Delta^*\Psi \sim \frac{F_0^2}{\mu_0 R^2}|\nabla^\perp\Psi|^2,
\end{gathered}
\end{equation*}
where we have assumed that the transient values of \(p\) and \(\Psi\) persist at the same order of magnitude as the equilibrium values \(p_0\) and \(\Psi_0\). As is usual in the tokamak limit, we have \(F_0 = RB_v = \const\). Comparing the time derivatives, we see that \(\partial_t [B^2/(2\mu_0)] \div \partial_t p \sim 1\).

\section{Reduced MHD}
\label{sec:rmhd}

Although there are many approaches to reduced MHD, which often involve an expansion with respect to the inverse aspect ratio, the common goal of all these models is the elimination of fast magnetosonic waves \cite{franck2015energy,strauss1997reduced,strauss1976nonlinear,strauss1977dynamics,strauss1980stellarator,kruger1998generalized}. By eliminating the fastest propagating waves, we decrease the maximum velocity in the system, thus increasing the maximum time step allowed by the Courant condition in numerical simulations with explicit time integration. When implicit methods are used, the Courant condition is no longer a hard limit, however using time steps that are large compared to the shortest time scale can lead to particularly stiff matrix systems and poor accuracy \cite{jardin2012multiple,kruger1998generalized}.

In this paper, we adopt an ansatz approach to reduced MHD, which does not rely on a large aspect ratio, does not assume an ordering and, in fact, does not require a toroidal geometry at all. While the assumption \eqref{eq:assmp} is identical to the first assumption of the ordering in \cite{strauss1997reduced}, we do not \emph{a priori} introduce further assumptions on the magnitudes of the other hydromagnetic variables. We will, however, derive further conditions from the equations themselves, see section \ref{sec:mcon}. In appendix \ref{sec:ordering}, we will show an alternative ordering-based approach which also does not assume a toroidal geometry. As long as assumption \eqref{eq:assmp} is met, i.e. in the presence of a strong guiding field, we can eliminate fast magnetosonic waves by setting \(\zeta = 0\). In addition, since the first two terms of the velocity expression \eqref{eq:vfield} do not compress the magnetic field, we can also set \(\Omega = 0\), further simplifying the equations. However, since \(\vec B_0 \approx \nabla\chi\) is only an approximation, the wave equations in section \ref{sec:waves} will only hold approximately. Thus, some residual effects of the fast magnetosonic wave can remain even after setting \(\zeta = 0\). Care must be taken when using an explicit time stepping method, as these residual effects can prevent the usage of time steps larger than the fast magnetosonic time scale, for this reason it is preferable to use implicit schemes as is done in modern codes. When implicit time stepping is used, the residual effects amount to large eigenvalues in the resulting matrices, which can lead to convergence issues with iterative solvers. These problems can usually be mitigated by applying a preconditioner \cite{franck2015energy,jardin2012multiple}.

Having eliminated two of the dependent variables, we can also drop the corresponding equations. In such a manner, we obtain the following set of reduced MHD equations:
\begin{widetext}
\begin{equation}
\label{eq:rrhoeq}
\tderiv{\rho} = - B_v\left[\frac{\rho}{B_v^2},\Phi\right] - B_v\llderiv(\rho v_\parallel) - B_v[\rho v_\parallel,\Psi] + P,
\end{equation}
\begin{equation}
\label{eq:reneq}
\begin{aligned}
\tderiv{}\left(\frac{\rho v^2}{2} + \frac{p}{\gamma-1} + \frac{B^2}{2\mu_0}\right) &= - B_v\left[\frac{1}{B_v^2}\left(\frac{\rho v^2}{2} + \frac{\gamma p}{\gamma-1} + \frac{B^2}{\mu_0}\right),\Phi\right] - B_v\llderiv\left[v_\parallel\left(\frac{\rho v^2}{2} + \frac{\gamma p}{\gamma-1} + \frac{B^2}{\mu_0}\right) - \frac{\vec v\cdot\vec B}{\mu_0}\right] \\
&- B_v\left[v_\parallel\left(\frac{\rho v^2}{2} + \frac{\gamma p}{\gamma-1} + \frac{B^2}{\mu_0}\right) - \frac{\vec v\cdot\vec B}{\mu_0},\Psi\right] - \frac{1}{\mu_0}\nabla\cdot(\eta\vec j\times\vec B) \\
&+ \nabla\cdot\Bigg[\frac{p}{\gamma-1}\frac{D_\perp}{\rho}\left(\nabla\rho-\frac{\vec B}{B^2}(B_v\llderiv\rho + B_v[\rho,\Psi])\right) \\
&+ \frac{\kappa_\perp}{R}\nabla\left(\frac{p}{\rho}\right) + \frac{(\kappa_\parallel-\kappa_\perp)\vec B}{RB^2}\left(B_v\llderiv\left(\frac{p}{\rho}\right) + B_v\left[\frac{p}{\rho},\Psi\right]\right)\Bigg] + S_e - \frac{v^2}{2}P
\end{aligned}
\end{equation}
where we now have
\begin{equation}
\label{eq:redexprs}
\begin{gathered}
P = \nabla\cdot\left[D_\perp\nabla\rho - \frac{D_\perp\vec B}{B^2}(B_v\llderiv\rho + B_v[\rho,\Psi])\right] + S_\rho, \\
\vec B = \nabla\chi + \nabla\Psi\times\nabla\chi, \\
\vec v = \frac{\nabla\Phi\times\nabla\chi}{B_v^2} + v_\parallel\vec B, \\
v^2 = \frac{1}{B_v^2}(\Phi,\Phi) + 2v_\parallel(\Phi,\Psi) + v_\parallel^2 B_v^2 + v_\parallel^2 B_v^2(\Psi,\Psi), \\
\vec v\cdot\vec B = (\Phi,\Psi) + v_\parallel B_v^2 + v_\parallel B_v^2(\Psi,\Psi), \\
B^2 = B_v^2 + B_v^2(\Psi,\Psi), \\
\vec j = \frac{1}{\mu_0}[-\nabla\chi\Delta\Psi + B_v\llderiv\nabla\Psi - (\nabla\Psi\cdot\nabla)\nabla\chi], \\
\vec\omega = \nabla\times\vec v = -\nabla\chi\nabla\cdot\left(\frac{\nabla\Phi}{B_v^2}\right) + B_v\llderiv\frac{\nabla\Phi}{B_v^2} - \frac{1}{B_v^2}(\nabla\Phi\cdot\nabla)\nabla\chi + \nabla v_\parallel\times\vec B + v_\parallel\vec j.
\end{gathered}
\end{equation}
Reduction of the flux and momentum equations results in the following three equations:
\begin{equation}
\label{eq:rmfeq}
\left[\psi_v,\tderiv{\Psi}\right] = \left[\frac{[\Psi,\Phi]-\llderiv\Phi}{B_v},\psi_v\right] + \frac{1}{B_v}\nabla\cdot(\eta\nabla\psi_v\times\vec j)
\end{equation}
\begin{equation}
\label{eq:rphieq}
\begin{aligned}
&\pLap\tderiv{\Phi} + \left(v_\parallel B_v^2,\tderiv{\Psi}\right) + v_\parallel B_v^2\pLap\tderiv{\Psi} = \nabla\cdot\Bigg[\frac{B_v\llderiv v^2}{2}\pgrad\Psi + \omega_\chi(\nabla\Phi\times\nabla\chi + B_v^2 v_\parallel\nabla\Psi\times\nabla\chi) - v_\chi B_v^2\vec\omega^\perp \\
&+ \frac{B^2}{\rho}\vec j^\perp - \frac{j_\chi B_v^2}{\rho}\nabla\Psi\times\nabla\chi - \frac{P}{\rho}\pgrad\Phi\Bigg] + B_v\left[\frac{1}{\rho},p\right] + \left(\frac{B_v\llderiv p}{\rho},\Psi\right) + \frac{B_v\llderiv p}{\rho}\pLap\Psi + \nu\Delta\pLap\Phi
\end{aligned}
\end{equation}
\begin{equation}
\label{eq:rvpeq}
\begin{aligned}
&\tderiv{v_\parallel} = -\frac{v_\parallel}{\rho}P + \frac{1}{B_v}\Bigg[\frac{\vec\omega\cdot\pgrad\Phi}{B_v} + B_v v_\parallel\vec\omega\cdot\pgrad\Psi - \frac{\llderiv v^2}{2} - \frac{B_v\vec j\cdot\pgrad\Psi - \llderiv p}{\rho}\Bigg] + \nu\Delta v_\parallel,
\end{aligned}
\end{equation}
\end{widetext}

A further simplification would be to set \(v_\parallel = 0\), reducing the number of unknowns to four at the expense of field-aligned flows. Similar reduced models for tokamaks that include field-aligned flows, as well as variants which only allow flow perpendicular to the background field, are used in the JOREK and M3D-$C^1$ codes \cite{huysmans2007mhd,breslau2009some}.

As a final remark, we show that in the reduced ideal case the scalar function \(\Phi\) introduced in the velocity expression \eqref{eq:vfield} corresponds, up to an additive constant, to the electric potential taken with the opposite sign. Faraday's law in potential form states
\begin{equation}
\tderiv{\vec A} = -\vec E - \nabla V,
\end{equation}
where \(V\) is the electric potential and \(\vec A\) is the magnetic vector potential. Using the ideal Ohm's law \(\vec E = -\vec v\times\vec B\) with the reduced expressions for \(\vec v\) and \(\vec B\) from \eqref{eq:redexprs} and the expression \eqref{eq:vpot} for \(\vec A\) with \(\Omega = 0\) (where we have dropped the vacuum field vector potential due to its static nature), we obtain
\begin{equation}
\tderiv{\Psi}\nabla\chi = -\pgrad\Phi - \frac{\nabla\chi}{B_v}[\Psi,\Phi] - \nabla V.
\end{equation}
Taking just the components perpendicular to the vacuum field, we get \(\pgrad\Phi = -\pgrad V\), which requires \({\Phi = -V + c(\chi)}\), where \(c(\chi)\) is an arbitrary function. The component along the vacuum field is essentially an evolution equation for \(\Psi\):
\begin{equation}
\tderiv{\Psi}\nabla\chi = (\llderiv\Phi-c'-[\Psi,\Phi])\frac{\nabla\chi}{B_v},
\end{equation}
where we have replaced \(V\) with \(-\Phi + c(\chi)\). To show that this equation is consistent with equation \eqref{eq:rmfeq}, we take the curl, project it on \(\nabla\psi_v\) and divide by \(B_v\), obtaining
\begin{equation}
\left[\psi_v,\tderiv{\Psi}\right] = \left[\frac{[\Psi,\Phi]-c'-\llderiv\Phi}{B_v},\psi_v\right],
\end{equation}
which, if \(c' = 0\), is exactly the ideal version of equation \eqref{eq:rmfeq}. Thus, \(c\) is an additive constant.

\section{Conservation properties}
\label{sec:conservation}

In this section we consider sources of error and validity conditions for the reduced MHD approximation by looking at the components of the MHD equations \eqref{eq:vrmhd} that are dropped in the reduction. Since all of the MHD equations \eqref{eq:vrmhd}, except for the induction equation, are local conservation laws, any error introduced by the reduction amounts to a non-conservation of the corresponding quantity. For the induction equation, which allows for non-conservation of flux even when it is satisfied exactly, we will consider both the physical non-conservation of flux due to resistivity as well as the reduction error.

The conservation of mass and energy is exact, due to the fact that the continuity and energy equations are used directly to evolve density and pressure, with non-conservation being only due to the terms \(S_\rho\) and \(S_e\), which correspond to physically meaningful sources. On the other hand, momentum is not conserved due to equation \eqref{eq:zetaeq} being discarded after the reduction. Indeed, if one were to set \(\zeta = 0\) and attempt to retain equation \eqref{eq:zetaeq}, one would be left with an overconstrained system, with more equations than unknowns. For the same reason the second equation in \eqref{eq:mfeqns} is discarded. Unlike momentum, flux is physically not conserved due to resistivity, and the reduction leads to errors in the rate of change of flux.

\subsection{Non-conservation of flux}

We follow the same general procedure to show non-conservation of flux due to finite resistivity as Freidberg \cite{freidberg2014ideal} does to show conservation in the ideal case. Magnetic flux through an arbitrary surface \(S(t)\) is defined as
\begin{equation}
\psi = \int\limits_{S(t)}\vec B\cdot d\vec S,
\end{equation}
where the surface \(S(t)\) is advected with the plasma, hence its dependence on time. Taking the time derivative, and then applying the induction equation \eqref{eq:vrmhd} and Stokes' theorem, we obtain
\begin{equation}
\frac{d\psi}{dt} = \int\limits_{S(t)}\tderiv{\vec B}\cdot d\vec S + \oint\limits_{\partial S(t)}\vec B\cdot(\vec v\times d\vec l) = -\oint\limits_{\partial S(t)}\eta\vec j\cdot d\vec l,
\end{equation}
where \(\partial S(t)\) is the loop enclosing \(S(t)\). This is the non-conservation of flux due to resistivity, which is present in the full MHD model, and, as expected, is locally proportional to the resistivity. Clearly, the resistive term in the induction equation \eqref{eq:vrmhd} is responsible for this non-conservation. Interestingly enough, when we apply the reduction, only the resistive term is left in the second equation in \eqref{eq:mfeqns}. Indeed, setting \(\Omega = \zeta = 0\) and multiplying by \(B_v\), the second equation in \eqref{eq:mfeqns} becomes:
\begin{equation*}
0 = \nabla\cdot(\eta\nabla\chi\times\vec j),
\end{equation*}
which is satisfied when \(\eta = 0\). Thus, in the ideal case, both equations \eqref{eq:mfeqns} can be satisfied even after a reduction, which means that the induction equation \eqref{eq:vrmhd} will be satisfied in reduced ideal MHD. As such, when \(\eta = 0\), flux will be conserved and magnetic field lines will be frozen into the plasma even in the reduced MHD model. For nonzero resistivity, the second equation in \eqref{eq:mfeqns} cannot be satisfied and must be dropped. This amounts to neglecting a component of the resistive term in the induction equation and, depending on the relative orientations of \(\vec j^\perp\) and the loop \(\partial S(t)\), underestimating or overestimating \(\partial\psi/\partial t\). Since the term in the above equation is also locally proportional to the perpendicular components of the current, we need \(|j^\parallel| \gg |\vec j^\perp|\) in order for \(\Omega = \zeta = 0\) to be a valid approximation. In other words, \(|\vec j^\perp|/j\) can serve as an order of magnitude estimate of the relative reduction error in \(\partial\psi/\partial t\). As we will see below, the perpendicular components of the current, which arise due to nonzero parallel derivatives of \(\Psi\) and components of the metric tensor of the vacuum field-aligned coordinate system, also contribute to momentum conservation error.

\subsection{Non-conservation of momentum}
\label{sec:mcon}

The action of the first projection operator \eqref{eq:projop} on a vector \(\vec s\) can be written as \(\nabla\chi\cdot\nabla\times[\nabla\chi\times(\vec e_\chi\times\vec s)] = \nabla\chi\cdot\nabla\times(\vec e_\chi s^\chi - \vec s)\). Thus, equation \eqref{eq:rphieq} is the contravariant \(\chi\) component of vorticity-type equation, which we will refer to as the reduced vorticity equation. If all three components of this reduced vorticity equation were satisfied (which, in general, is not possible as the system of equations would be overconstrained), then the original Navier-Stokes equation would also be satisfied and momentum would be conserved exactly. We can therefore estimate the magnitude of momentum conservation error by considering the components of the vorticity-type equation perpendicular to \(\nabla\chi\). This vorticity-type equation can be written as
\begin{widetext}
\begin{equation}
\label{eq:vort}
\begin{aligned}
&\tderiv{\st{\vec\omega}} + v_\parallel\tderiv{\vec j} + \nabla v_\parallel\times\tderiv{\vec B} - \frac{\vec j}{B_v}\llderiv v^2 - \nabla\left(\frac{\llderiv v^2}{B_v}\right)\times\vec B + \nabla\times(\vec\omega\times\vec v) - \frac{\vec j}{B_v^2}\nabla\chi\cdot(\vec\omega\times\vec v) \\
&- \nabla\left[\frac{\nabla\chi\cdot(\vec\omega\times\vec v)}{B_v^2}\right]\times\vec B + \st{\vec\omega}~\frac{P}{\rho} + \nabla\left(\frac{P}{\rho}\right)\times\st{\vec v} = \frac{\vec B}{\rho^2}(\vec j\cdot\nabla\rho) - \frac{\vec j}{\rho^2}(\vec B\cdot\nabla\rho) + \frac{1}{\rho}(\vec B\cdot\nabla)\vec j - \frac{1}{\rho}(\vec j\cdot\nabla)\vec B \\
&- \frac{\vec j}{\rho B_v^2}\nabla\chi\cdot(\vec j\times\vec B) - \nabla\left[\frac{\nabla\chi\cdot(\vec j\times\vec B)}{\rho B_v^2}\right]\times\vec B + \frac{1}{\rho^2}\nabla\rho\times\nabla p + \frac{\vec j}{\rho B_v}\llderiv p + \nabla\left(\frac{\llderiv p}{\rho B_v}\right)\times\vec B,
\end{aligned}
\end{equation}
\end{widetext}
where we have introduced the reduced velocity \(\st{\vec v} = -\nabla\chi\times(\vec e_\chi\times\vec v) = \nabla\Phi\times\nabla\chi/B_v^2\) and the reduced vorticity \(\st{\vec\omega} = \nabla\times\st{\vec v}\). The viscosity term is not considered here since we have not done a proper derivation of it but simply added a generic term after the fact. If the components of this equation perpendicular to \(\nabla\chi\) are identically zero, then there is no approximation in the velocity reduction as nothing is being neglected, and momentum is still conserved. The most general case in which the perpendicular components are zero is the following:
\begin{equation}
\label{eq:exact}
\llderiv u = 0, \quad u\in\{g^{ik}, \Phi, \Psi, v_\parallel, p, \rho, P\}
\end{equation}
where \(g^{ik}\) are the components of the metric tensor of the vacuum field-aligned coordinate system. As can be shown by a simple calculation, in this case both \(\st{\vec\omega}\) and \(\vec j\) will be directed strictly along \(\nabla\chi\). If we allow either the metric tensor, \(\Phi\) or \(\Psi\) to vary along \(\nabla\chi\), the same calculation will show that \(\st{\vec\omega}\) has nonzero perpendicular components. This will cause \(\partial\st{\vec\omega}/\partial t\) to have nonzero perpendicular components, which cannot be canceled by any other terms since there are no more time derivatives involving \(\Phi\) in the equation. Similarly, if we let any of the other quantities vary along \(\nabla\chi\), the last term and the seventh term on the RHS (pressure), the first term and the seventh term on the RHS (density), the last term on the LHS (\(P\)) and third term on the LHS (\(v_\parallel\)) will be nonzero and will not be canceled by any other terms. If the conditions \eqref{eq:exact} are met, then only the fourth and sixth terms on the LHS are nonzero. As can be shown by a simple expansion of the sixth term:
\begin{equation*}
[\nabla\times(\vec\omega\times\vec v)]^\perp - \left[\nabla\left[\frac{\nabla\chi\cdot(\vec\omega\times\vec v)}{B_v^2}\right]\times\vec B\right]^\perp \equiv 0.
\end{equation*}
A major simplification comes from the fact that, due to \(\llderiv g^{\psi_v\beta_v} = 0\), the Clebsch-type coordinate system aligned to the vacuum field can be made orthogonal \cite{salat2000conditions}. In addition, the conditions \eqref{eq:exact} also require that the lengths of the basis vectors do not vary along \(\nabla\chi\), which forces all of the co- and contravariant components of \(\vec v\) and \(\vec\omega\) to be constant along vacuum field lines.

In the general case, when the conditions \eqref{eq:exact} are no longer satisfied, the velocity as given by \eqref{eq:redexprs} is no longer an exact solution to the Navier-Stokes equation, and momentum is not conserved exactly. Nevertheless, as long as we have \(|\llderiv u| \ll |\pgrad u|\), the approximation \(\nabla u \approx \pgrad u\) is valid and errors introduced by the reduction should be small. The smallness of the parallel derivative is, in most cases, a reasonable assumption, and is included in most orderings \cite{strauss1997reduced,strauss1976nonlinear,strauss1977dynamics,strauss1980stellarator,kruger1998generalized}. A notable example when this assumption is not valid is pellet injection \cite{futatani2014non}, when, at the tip of simulated pellet, the local gradients of density, pressure and \(P\) can be comparable in the parallel and perpendicular directions. However, during pellet ablation, the density and pressure perturbations will quickly equilibrate along the total magnetic field, and if the direction of the total field is not too different from the direction of the vacuum field, as implied by \eqref{eq:assmp}, the parallel derivatives will return to being small even in such a scenario.

The metric tensor is the only quantity in \eqref{eq:exact} that is determined solely by the vacuum field and is not affected by the dynamics of the system. The assertions \eqref{eq:exact} imply that the vacuum field has zero local shear everywhere \cite{salat2000conditions}, and that its strength does not change along field lines. As mentioned previously in section \ref{sec:rep}, we have a degree of control over what we choose to be the "vacuum field", as long as the assumption \eqref{eq:assmp} is valid, otherwise the MHD waves will not separate in the velocity representation \eqref{eq:vfield}. Of course, \(\chi\) must always satisfy the Laplace equation in order for \(\nabla\chi\) to be a valid magnetic field. In the case of a tokamak, choosing \(\chi = F_0\phi\), where \(F_0\) is a constant and \(\phi\) is the toroidal angle, will satisfy \(\llderiv g^{ik} = 0\) exactly. The part of the vacuum field created by the poloidal coils will then be grouped with the induced field. If a two-dimensional, completely axisymmetric tokamak simulation is run, then the conditions \eqref{eq:exact} will be satisfied exactly. In the more general case of an arbitrary three-dimensional magnetic configuration, as in a stellarator, it may not be possible to choose \(\chi\) so that \(\llderiv g^{ik} = 0\) is satisfied exactly while fulfilling the assumption \eqref{eq:assmp} at the same time. The choice of \(\chi\) would then be a compromise between \(|\llderiv g^{ik}| \ll |\pgrad g^{ik}|\) and assumption \eqref{eq:assmp}, with the parallel derivative of the metric tensor contributing to the error, which should still be small if the perpendicular derivatives are sufficiently large.

\section{Conclusion}

In the present article, a hierarchy of models suitable for stellarator geometies with excellent conservation properties was derived. We introduced representations that consist of a background vacuum field, a field line bending term and a field compression term for the magnetic field, and an \(\vec E\times\vec B\) term, a field-aligned flow term and a fluid compression term for the velocity. We also showed that any arbitrary magnetic and velocity fields can be expressed in this form. Thus, when we insert the representations into the viscoresistive MHD equations and apply appropriate projection operators to Faraday’s law and the Navier-Stokes equation, obtaining a system of scalar equations that is closed by the continuity and energy equations, the scalar equations are identical to the original full MHD equations in the inviscid case.

Importantly, we showed that, if the background vacuum field is stronger than the bending and compression terms, and if the \(\beta\) is sufficiently low, MHD waves are approximately separated in the velocity represenation, with Alfven waves contained in the \(\vec E\times\vec B\) term, slow magnetosonic waves in the field-aligned flow term and fast magnetosonic waves in the fluid compression term. Thus, by setting the fluid compression term to zero, we eliminated fast magnetosonic waves, obtaining a reduced MHD model. We also showed that the \(\vec E\times\vec B\) and field-aligned flow terms do not compress the magnetic field, which allows us to set the field compression term in the magnetic field representation to zero within the same reduced model. As an optional further reduction, we also considered a model where the field-aligned flow term is set to zero. This is similar to the approach followed by Breslau \emph{et al} \cite{breslau2009some} and Izzo \emph{et al} \cite{izzo1985reduced} for tokamaks.

Finally, by considering the terms that were neglected in the reduction, we showed that there is no approximation associated with the reduction if the background vacuum field is shear-free, all the unknown scalar fields do not vary in the direction of the background field, and the mass diffusivity \(D_\perp\) and density source \(S_\rho\) also do not vary in the direction of the background field. When this is not the case, the reduction leads to violations of the conservation of momentum and errors in the evolution of magnetic flux. Therefore, the reduction approximation is valid as long as the shear in the background vacuum field is low and parallel derivatives are small compared to perpendicular derivatives, which is often the case since parallel dynamics are much faster than perpendicular dynamics.

Although we derived these equations with the intention of eventually implementing them in the JOREK code, we deliberately made no assumptions about the underlying geometry of the problem. It should therefore be possible to apply these equations not only to toroidal fusion machines, such as tokamaks and stellarators, but also to non-toroidal and open field line configurations, which are often encountered in astrophysics, provided a strong guide field.

It remains an open question of how well various instabilities are represented in the two reduced MHD models presented here. We are currently working on implementing the reduced and full models in JOREK, after which we will benchmark the simulation results of the reduced models against the full model. We also plan to study the (1,1) internal kink mode in a tokamak analytically within the context of our reduced models.

\begin{acknowledgments}
The authors thank Omar Maj, Florian Hindenlang, Per Helander, Boniface Nkonga, Guido Huijsmans, Erika Strumberger, Alessandro Biancalani and Rohan Ramasamy for fruitful discussions.

This work has been carried out within the framework of the EUROfusion Consortium and has received funding from the Euratom research and training program 2014-2018 and 2019-2020 under grant agreement No 633053. The views and opinions expressed herein do not necessarily reflect those of the European Commission.
\end{acknowledgments}

\appendix
\section{Ordering}
\label{sec:ordering}

Reduced MHD equations similar, though not identical, to those considered in section \ref{sec:rmhd} can be obtained from the full MHD equations for \(\rho\), \(p\), \(\Phi\), \(\Psi\) and \(v_\parallel\) in section \ref{sec:derivation} using an ordering approach. We introduce a small parameter \(\lambda \equiv L_\perp/L_\parallel\), where \(L_\perp\) and \(L_\parallel\) are the perpendicular and parallel length scales, respectively. Then clearly the spatial derivatives must be ordered as follows:
\begin{equation}
\label{eq:sder}
|\llderiv|\sim\lambda|\pgrad|.
\end{equation}
Since we intend to eliminate fast magnetosonic waves, the fastest remaining waves will be Alfv\'{e}n waves, which travel along field lines. The shortest time scale will then be the Alfv\'{e}n time \(\tau_A \equiv L_\parallel/c_A\), and the partial time derivative will be ordered as:
\begin{equation}
\label{eq:tder}
\left|\tderiv{}\right| \sim \frac{1}{\tau_A} \equiv \frac{c_A}{L_\parallel},
\end{equation}
where \(c_A^2 \equiv B_s^2/(\mu_0\rho_s)\), and \(B_s\) and \(\rho_s\) are typical reference values for the magnetic field and density in the system. In addition, we make the following assumptions on the relative orders of the other quantities. The terms in the magnetic field represenation \eqref{eq:mfield} are ordered as:
\begin{equation}
\label{eq:b2}
\frac{|\nabla\Psi\times\nabla\chi|}{|\nabla\chi|} \sim |\pgrad\Psi| \sim \lambda,
\end{equation}
and
\begin{equation}
\label{eq:b3}
\frac{|\nabla\Omega\times\nabla\psi_v|}{|\nabla\chi|} \sim \frac{F_v}{B_v}|\nabla\Omega| \sim \lambda^3.
\end{equation}
For the terms in the velocity represenation \eqref{eq:vfield} we require:
\begin{equation}
\label{v2}
|v_\parallel\vec B|\div\frac{|\nabla\Phi\times\nabla\chi|}{B_v^2} \sim B_v^2\frac{|v_\parallel|}{|\pgrad\Phi|} \sim 1,
\end{equation}
and
\begin{equation}
\label{eq:v3}
|\pgrad\zeta|\div\frac{|\nabla\Phi\times\nabla\chi|}{B_v^2} \sim B_v\frac{|\pgrad\zeta|}{|\pgrad\Phi|} \sim \lambda.
\end{equation}
Finally, we assume that the partial and convective terms in the material derivative are of the same order:
\begin{equation}
\label{eq:conv}
\left|\tderiv{}\right| \div |\vec v\cdot\nabla| \sim \left|\tderiv{}\right| \div \frac{|\pgrad\Phi||\pgrad|}{B_v} \sim 1.
\end{equation}
We also assume that
\begin{equation}
\label{eq:auxa}
\begin{gathered}
\beta \sim \lambda^2,\qquad \kappa_\perp \sim \lambda\kappa_\parallel,\qquad \mathrm{Re}\sim\mathrm{Re_m}\sim 1, \\
P\sim \partial\rho/\partial t,\qquad S_e \sim \partial\mathcal{E}/\partial t
\end{gathered}
\end{equation}
Here, \(\mathcal{E} = \rho v^2/2 + p/(\gamma-1) + B^2/(2\mu_0)\) is the total energy.


If we normalize all lengths by \(L_\perp\), velocities by \(c_A\), densities by \(\rho_s\), thermal conductivities by \(\kappa_s\), \(\nabla\psi_v\) by \(F_s\), and magnetic fields by \(B_s \equiv c_A\sqrt{\mu_0\rho_s}\), deriving all other normalization factors from simple combinations of these, we obtain the following ordering for the normalized quantities:
\begin{itemize}
\item \(O(1):~\widetilde{\rho},~\widetilde{B_v},~\widetilde{F_v},~\widetilde{\kappa_\parallel},~\widetilde{\pgrad}\)
\item \(O(\lambda):~\widetilde{\Psi},~\widetilde{\Phi},~\widetilde{v_\parallel},~\widetilde{\eta},~\widetilde{\nu},~\widetilde{D_\perp},~\widetilde{\kappa_\perp},~\widetilde{S_\rho},~\widetilde{\llderiv},~\partial/\partial\widetilde{t}\)
\item \(O(\lambda^2):~\widetilde{\zeta},~\widetilde{p}\)
\item \(O(\lambda^3):~\widetilde{\Omega},~\widetilde{S_e}\)
\end{itemize}
where, for each quantity \(u\), \(\widetilde{u} = u/u_s\) and \(u_s\) is the normalization factor. The scales \(L_\perp\), \(L_\parallel\), \(\rho_s\), \(\kappa_s\), \(B_s\) and \(F_s\) were chosen so that \(L_\perp|\pgrad| \sim L_\parallel|\llderiv| \sim \widetilde{\rho} \sim \widetilde{\kappa_\parallel} \sim \widetilde{B} \sim \widetilde{F_v} \sim 1\).

We can now apply our ordering to the equations in section \ref{sec:derivation} either directly using relations (\ref{eq:sder}--\ref{eq:auxa}) or by first normalizing the equations and then using the ordering for the normalized quantities above. In any case, we get the same result, and if we drop the tildes from the normalized quantities, the resulting equations will be visually identical to the non-normalized equations.

Applying the ordering to equation \eqref{eq:rhoeq}, and keeping only the lowest order terms (\(O(\lambda)\) in this case), we obtain:
\begin{equation}
\label{eq:orhoeq}
\tderiv{\rho} = -B_v\left[\frac{\rho}{B_v^2},\Phi\right] + P,
\end{equation}
where \(P = \nabla\cdot(D_\perp\nabla\rho) + S_\rho\). There are two differences this equation has with equation \eqref{eq:rrhoeq}. First, the \(v_\parallel\) terms are eliminated by the ordering, which means that advection is dominated by the \(\vec E\times\vec B\) flow. This is a consequence of constraining the field-aligned flow to be of the same order of magnitude as the \(\vec E\times\vec B\) flow, as, in the lowest order, field-aligned flow is in the \(\nabla\chi\) direction, which produces a parallel derivative. The second difference is that the diffusion in the \(P\) term is no longer anisotropic. This can be justified by considering that diffusive transport is negligible compared to the transport due to \(v_\parallel\), so failing to subtract out diffusion along \(\vec B\) makes no difference.

When applying the ordering to equation \eqref{eq:eneq}, we note that while \(\mathcal{E}\) itself has order \(O(1)\) due to the \(B_v^2/(2\mu_0)\) term, \(\partial\mathcal{E}/\partial t\) will be \(O(\lambda^3)\) and \(\pgrad(\mathcal{E}/B_v^2)\) will be \(O(\lambda^2)\). In addition, in the expression \(v_\parallel\mathcal{E}-\vec v\cdot\vec B/\mu_0\), the \(O(\lambda)\) term \(v_\parallel B_v^2/\mu_0\) is cancelled by a similar term in \(\vec v\cdot\vec B\). Keeping just the lowest order terms (\(O(\lambda^2)\) this time), the following is obtained after some simplifications:
\begin{equation}
\label{eq:ozetaeq}
2(B_v,\zeta) + B_v\pLap\zeta = 0.
\end{equation}
We will treat this as the governing equation for \(\zeta\). We do not need to actually solve it since we are only interested in the first two terms of \(\vec v\) and \(\zeta\) does not appear in the governing equations of any quantities that we are interested in. It should be pointed out that the perpendicular components of \(\vec j\) are all \(O(\lambda^2)\) or higher. In particular, if we expand the \((\nabla\Psi\cdot\nabla)\nabla\chi\) term and evaluate the Christoffel symbols, we see that its perpendicular components are also \(O(\lambda^2)\). Taking the next lowest order (\(O(\lambda^3)\)) terms in equation \eqref{eq:eneq}, we have
\begin{equation}
\label{eq:oeneq}
\begin{aligned}
\tderiv{\mathcal{E}} &= - B_v\left[\frac{\mathcal{E} + p}{B_v^2},\Phi\right] + \frac{B_v}{\mu_0}\llderiv(\Phi,\Psi) + \frac{B_v}{\mu_0}[(\Phi,\Psi),\Psi] \\
&- \frac{1}{\mu_0}\nabla\cdot[\eta\vec j\times(\nabla\chi + \nabla\Psi\times\nabla\chi)] + \nabla\cdot\Bigg[\frac{\kappa_\perp}{R}\nabla\left(\frac{p}{\rho}\right) \\
&+ \frac{p}{\gamma-1}\frac{D_\perp}{\rho}\nabla\rho + \frac{\kappa_\parallel\nabla\chi}{RB_v}\left(\llderiv\left(\frac{p}{\rho}\right) + \left[\frac{p}{\rho},\Psi\right]\right)\Bigg] \\
&+ S_e - \frac{v^2}{2}P,
\end{aligned}
\end{equation}
where the expressions for \(v^2\), \(B^2\) and \(\vec j\) match those given in \eqref{eq:redexprs}. Comparing this to equation \eqref{eq:reneq}, we se that, again, the \(\vec E\times\vec B\) flow dominates advection and field-aligned advection is not present at this order. The subtracting out of the component of \(\kappa_\perp\) heat transport in the \(\vec B\) direction is neglected since \(\kappa_\parallel \gg \kappa_\perp\), similarly to our treatment of mass diffusion. Finally, in the parallel heat flow term, \(\vec B\) and \(B^2\) have been approximated by \(\nabla\chi\) and \(B_v^2\), respectively, neglecting the higher order differences.

We proceed to equation \eqref{eq:vmfeq}. The lowest order (\(O(\lambda^2)\)) terms are
\begin{equation}
\label{eq:ovmfeq}
\nabla\tderiv{\Psi}\times\nabla\chi = \nabla\times\left[\frac{\nabla\chi}{B_v}(\llderiv\Phi-[\Psi,\Phi]) + \nabla\zeta\times\nabla\chi - \eta\vec j\right].
\end{equation}
Projecting this equation on \(\nabla\chi\), we get exactly the same equation as \eqref{eq:ozetaeq}. Projecting on \(\nabla\psi_v\), we get
\begin{equation}
\label{eq:opsieq}
\left[\psi_v,\tderiv{\Psi}\right] = \left[\frac{[\Psi,\Phi]-\llderiv\Phi}{B_v},\psi_v\right] + \frac{\nabla\cdot(\eta\nabla\psi_v\times\vec j)}{B_v},
\end{equation}
which exactly matches equation \eqref{eq:rmfeq}. We do not need an equation for \(\Omega\) since \(\Omega\) does not appear in the governing equations for any of quantities that we are interested in.

Now consider equation \eqref{eq:phieq}. The lowest order terms are \(O(\lambda^2)\):
\begin{equation}
\label{eq:ophieq}
\begin{aligned}
\pLap\tderiv{\Phi} &= \nabla\cdot\Bigg[\frac{\omega^\chi}{B_v^2}(\nabla\Phi\times\nabla\chi + B_v^2 v_\parallel\nabla\Psi\times\nabla\chi) \\
&- v_\parallel B_v^2\vec\omega^\perp + \frac{B_v^2}{\rho}\vec j^\perp - \frac{j^\chi}{\rho}\nabla\Psi\times\nabla\chi - \frac{P}{\rho}\pgrad\Phi\Bigg] \\
&+ B_v\left[\frac{1}{\rho},p\right] + \nu\Delta\pLap\Phi.
\end{aligned}
\end{equation}
This result is equivalent to approximating \(\vec B\) by \(\nabla\chi\) in equation \eqref{eq:nsins}, which will cause the third term on the LHS to disappear, and then applying the projection operator, which is now \(\nabla\chi\cdot\nabla\times[B_v^{-2}\nabla\chi\times(\nabla\chi\times\) due to the approximation we made. Note that we have also approximated \(\vec B\) by \(\nabla\chi\) in the projections (\(s_\chi\) becomes \(s^\chi/B_v^2\) for any vector \(\vec s\)) and \(B^2\) by \(B_v^2\) in the coefficient in front of \(\vec j^\perp\).

The last equation that we consider is equation \eqref{eq:vpeq}. The lowest order terms are again \(O(\lambda^2)\):
\begin{equation}
\label{eq:ovpeq}
\tderiv{v_\parallel} = -\frac{v_\parallel}{\rho}P + \frac{\vec\omega\cdot\pgrad\Phi}{B_v^2} + \nu\Delta v_\parallel.
\end{equation}
Just as equation \eqref{eq:ophieq}, this equation corresponds to approximating \(\vec B\) by \(\nabla\chi\) in equation \eqref{eq:nsins} and applying the projection operator \(\nabla\chi\cdot\). In addition, both the hydrodynamic pressure and ram pressure are neglected.

Finally, we must simply drop equation \eqref{eq:zetaeq} to avoid having an overconstrained system. If we attempt to keep it, \(\zeta\) and \(\Omega\) will drop from the equation in the lowest order, leaving us with an equation involving just \(\Psi\), \(\Phi\), \(v_\parallel\) and \(p\), and overconstraining these variables. Our decision is in line with what is generally done in ordering approaches \cite{strauss1997reduced,strauss1976nonlinear,strauss1977dynamics,strauss1980stellarator,kruger1998generalized}, where one considers the parallel projection of the vorticity equation and the parallel projection of the Navier-Stokes equation, but not the divergence of the Navier-Stokes equation.

We have thus shown that a similar, though much simpler set of reduced MHD equations can be derived using an ordering. However, more assumptions are needed for an ordering than for a consistent ansatz apprach, and these extra assumptions are what allows us to obtain simpler equations.

\bibliography{references}

\end{document}